\documentclass[conference,compsoc]{IEEEtran}
\usepackage{xcolor}
\usepackage{graphicx} 
\usepackage{subcaption}
\usepackage{adjustbox}
\usepackage{stfloats} 
\usepackage{tikz}
\usepackage{amsmath}
\usepackage{mathtools}
\usepackage{enumitem}
\usepackage{booktabs}
\usepackage[T1]{fontenc}
\setlist[itemize]{leftmargin=20pt}
\usetikzlibrary{fit}
\usetikzlibrary{calc}
\usetikzlibrary{shapes.misc, positioning}
\usetikzlibrary{automata,arrows,positioning,calc}
\usepackage{balance}
\usepackage{longtable}
\usepackage{pifont}
\usepackage{xcolor}
\usepackage{url}

\definecolor{lightgrey}{HTML}{d3d3d3}
\definecolor{LIGHTBLUE}{HTML}{06AED5}
\definecolor{RED}{HTML}{DA3E52}
\definecolor{YELLOW}{HTML}{F8BE57}
\definecolor{GREEN}{HTML}{02C39A}
\definecolor{DARKBLUE}{HTML}{05668D}
\definecolor{VIOLET}{HTML}{B118C8}
\definecolor{ORANGE}{HTML}{FF784F}
\definecolor{ROSE}{HTML}{B98389}
\definecolor{PINK}{HTML}{F2BAC9}
\definecolor{LIGHTGREEN}{HTML}{C1CC99}
\definecolor{DARKPURPLE}{HTML}{3c096c}
\definecolor{LIGHTCYAN}{HTML}{D9F7FA}
\newcommand\T[1]{\vspace{2pt} \textbf{#1} }

\newif\ifEditMode
\EditModetrue
\usepackage{balance}
\usepackage{supertabular}

\usepackage{aliases}
\usepackage{preamble}

\ifCLASSOPTIONcompsoc
  \usepackage[nocompress]{cite}
\else
  \usepackage{cite}
\fi
\ifCLASSINFOpdf
\else
\fi

\begin{document}
%
\title{Non-Atomic Arbitrage in Decentralized Finance}

\author{\IEEEauthorblockN{Lioba Heimbach}
\IEEEauthorblockA{ETH Zurich\\
Switzerland\\
hlioba@ethz.ch}
\and
\IEEEauthorblockN{Vabuk Pahari}
\IEEEauthorblockA{MPI-SWS\\ Germany\\
vpahari@mpi-sws.org}
\and
\IEEEauthorblockN{Eric Schertenleib}
\IEEEauthorblockA{unaffiliated\\
Switzerland\\
eric.schertenleib@gmail.com}}


%


\maketitle

\begin{abstract}
The prevalence of \textit{maximal extractable value (MEV)} in the Ethereum ecosystem has led to a characterization of the latter as a \textit{dark forest}. Studies of MEV have thus far largely been restricted to purely on-chain MEV, i.e., sandwich attacks, cyclic arbitrage, and liquidations. In this work, we shed light on the prevalence of non-atomic arbitrage on decentralized exchanges (DEXes) on the Ethereum blockchain. Importantly, non-atomic arbitrage exploits price differences between DEXes on the Ethereum blockchain as well as exchanges outside the Ethereum blockchain (i.e., centralized exchanges or DEXes on other blockchains). Thus, non-atomic arbitrage is a type of MEV  that involves actions on and off the Ethereum blockchain. 

In our study of non-atomic arbitrage, we uncover that more than a fourth of the volume on Ethereum's biggest five DEXes from the merge until 31 October 2023 can likely be attributed to this type of MEV. We further highlight that only eleven searchers are responsible for more than 80\% of the identified non-atomic arbitrage volume sitting at a staggering \$132~billion and draw a connection between the centralization of the block construction market and non-atomic arbitrage. Finally, we discuss the security implications of these high-value transactions that account for more than 10\% of Ethereum's total block value and outline possible mitigations.

\end{abstract}


%
\IEEEpeerreviewmaketitle

\section{Introduction}
The introduction of \textit{decentralized exchanges (DEXes)} has made the trading of cryptocurrencies a centerpiece of \textit{decentralized finance (DeFi)}. Today, a multitude of DEXes exist on several different chains. The prices on each of these DEXes are the product of previous transactions on the respective DEX and thus can differ from the prices quoted at a given time on centralized exchanges (CEX) or other DEXes. These differences represent an arbitrage opportunity, which a trader can exploit by buying the asset at the cheaper venue and selling it at the other. We refer to price differences between DEXes on the same chain as \textit{atomic arbitrage}, since these arbitrages can be carried out within the same transaction, whereas \textit{non-atomic arbitrage} stems from price discrepancies involving DEXes on two different chains or, presumably more common, centralized exchanges. The prior is known as cyclic arbitrage and has been extensively studied~\cite{WangCyclic2022}. We focus on the latter, which is naturally more challenging to analyze as it involves actions outside the respective blockchain. We note that while non-atomic arbitrage opportunities exist on various chains we focus on Ethereum, as it is home to the by far largest portion of DeFi applications~\cite{Defillama-chain-data}. 

Non-atomic arbitrage opportunities existed ever since the launch of DEXes, as these naturally arise when you have two markets quoting prices for the same assets. However, Ethereum's transition from \textit{Proof-of-Work (PoW)} to \textit{Proof-of-Stake (PoS)} in September 2022 marked a watershed moment, due to the changes in block building. On Ethereum PoS time is divided into slots with each slot being assigned to a single known \textit{validator}, i.e., the PoS equivalent of a miner, who is responsible for proposing a block and extending the chain. While validators are assigned a slot, around 90\% of blocks are no longer built by the validator themselves, but are instead outsourced through the novel \textit{proposer-builder separation (PBS)} to a small set of specialized builders~\cite{2022boost}. These builders bid to have their assembled block selected by the validator. The acquired block building right is valuable partly due to users paying \textit{tips} for block inclusion, but, more importantly, due to \textit{maximal extractable value (MEV)}, i.e., the value that can be extracted by strategically ordering, including or excluding transactions. Trades exploiting non-atomic arbitrage opportunities make up a part of MEV as arbitrageurs want to be the first to exploit this opportunity, and therefore are keen to be included at the top of the block~\cite{Gupta2023Centralizing}.

These changes have generally made it easier for sophisticated traders to extract value from non-atomic arbitrage opportunities. Firstly, whereas on Ethereum PoW the miner was a priori unknown, the advance knowledge of the validator is beneficial for the arbitrageurs as they can identify slots that have been assigned to validators open to external block builders, i.e., validators participating in PBS. Through PBS arbitrageurs have a scheme by which they can relay their transactions privately to the builder and do not risk being front-run in the \textit{mempool}, i.e., the public waiting area for transactions. Secondly, block builders are incentivized to maximize MEV, with some even specializing in non-atomic arbitrage~\cite{Gupta2023Centralizing}, institutionalizing it in block building.

In this work, we present the first in-depth analysis of non-atomic arbitrage on Ethereum DEXes to show that non-atomic arbitrage, a type of MEV, accounts for a significant proportion of the total DEX volume. We further identify alarming centralizing trends in the block building market as a direct consequence of non-atomic arbitrage.

\T{Contributions.} We summarize our main contributions as follows: 
\begin{itemize}[leftmargin=*]
    \item We develop a model to quantify the profits from non-atomic arbitrageurs.
    \item We perform the first in-depth measurement study of non-atomic arbitrage on DEXes and find that more than one-fourth of the volume on the five biggest DEXes on Ethereum is most likely non-atomic arbitrage.
    \item Our work finds that two searchers account for nearly half the non-atomic arbitrage volume and highlights that several builders operate (subsidized) integrated searchers (i.e., the builder and searcher are controlled by the same entity) doing non-atomic arbitrage.
    \item We show that non-atomic arbitrage is linked to times of high cryptocurrency price volatility and that during these times builders specializing in non-atomic arbitrage have higher chances of being selected -- centralizing the block construction market. 
    \item We discuss the implication of non-atomic arbitrage on the ecosystem and point towards possible mitigations. 
\end{itemize}

\section{Background}
In the following, we detail the relevant background of Ethereum PoS (cf. Section~\ref{sec:ETHPoS}), PBS (cf. Section~\ref{sec:PBS}), DEXes (cf. Section~\ref{sec:DEXes}), and MEV (cf. Section~\ref{sec:MEV}). 
\subsection{Ethereum Proof-of-Stake}\label{sec:ETHPoS}
Since the transition to PoS, Ethereum runs on two layers. The \textit{execution layer}, largely taken over from the former PoW protocol, is tasked with validating and executing transactions. Conversely, the \textit{consensus layer}, situated atop the \textit{beacon chain}, is dedicated to reaching consensus among validators, i.e., the PoS equivalent of miners. To become a validator, one must lock, i.e., \textit{stake}, 32~ETH in a specified smart contract. 

Time is divided into slots, periods of twelve-second length, on the Beacon chain. In every slot, a single validator is responsible for proposing a block, i.e., being a \textit{proposer}. Thus, there is a chance for a block to be proposed and for the blockchain to be extended in every slot. For building a block, a validator receives a fixed consensus layer block reward ($\approx 0.04$~ETH). Additionally, the validator also receives a variable execution layer reward for building the block. This variable reward stems from fees being paid for transactions to be included in the block. These fees can be relayed to the proposer either directly through transaction fees or through direct payment (i.e., \textit{coinbase transfers}). On average, these execution layer block rewards are around 0.12~ETH but vary significantly as we find. 

\subsection{Proposer Builder Separation}\label{sec:PBS}
PBS is a novel scheme that was designed for Ethereum PoS~\cite{2022pbs} and through which the majority (around 90\%~\cite{2022boost}) of blocks are proposed. With PBS the previous roles of the block proposer are separated into two tasks, namely block building and block proposing. In short, specialized block builders are tasked with building high-value blocks. Amongst these blocks built by the builders, the validator is expected to choose the block with the highest \textit{bid}, i.e., the amount the proposer would receive, and proposes the block to the network.

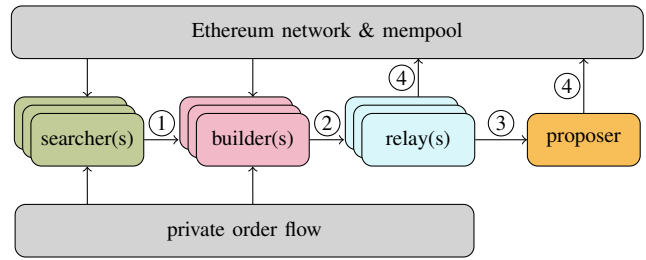
\begin{figure}[h]
    \centering
    \begin{tikzpicture}[scale=1.1]
  \node[draw, rounded corners, minimum width=8.4cm, minimum height=0.7cm,fill=lightgrey, font=\footnotesize] at (-3.1, 1.3)  {Ethereum network \& mempool};
  
  \node[draw, rounded corners, minimum width=1.5cm, minimum height=0.7cm, fill=YELLOW,font=\footnotesize] (proposer) at (0, 0)  {proposer};

  \node[draw, rounded corners, minimum width=1.5cm, minimum height=0.7cm,fill=LIGHTCYAN] at (-2.2, 0.2) {};
  \node[draw, rounded corners, minimum width=1.5cm, minimum height=0.7cm,fill=LIGHTCYAN] at (-2.1, 0.1) {};
  \node[draw, rounded corners, minimum width=1.5cm, minimum height=0.7cm,fill=LIGHTCYAN, font=\footnotesize] (relay) at (-2, 0)  {relay(s)};

  \node[draw, rounded corners, minimum width=1.5cm, minimum height=0.7cm,fill=PINK] at (-4.2, 0.2)  {};
  \node[draw, rounded corners, minimum width=1.5cm, minimum height=0.7cm,fill=PINK] at (-4.1, 0.1) {};
  \node[draw, rounded corners, minimum width=1.5cm, minimum height=0.7cm,fill=PINK, font=\footnotesize] at (-4, 0) (builder) {builder(s)};

  \node[draw, rounded corners, minimum width=1.5cm, minimum height=0.7cm,fill=LIGHTGREEN] at (-6.2, 0.2) {} ;
  \node[draw, rounded corners, minimum width=1.5cm, minimum height=0.7cm,fill=LIGHTGREEN] at (-6.1, 0.1) {};
  \node[draw, rounded corners, minimum width=1.5cm, minimum height=0.7cm,fill=LIGHTGREEN, font=\footnotesize] at (-6, 0) (searcher) {searcher(s)};
\node[draw, rounded corners, minimum width=6.1cm, minimum height=0.7cm,fill=lightgrey, font=\footnotesize] at (-4.1, -1.1) {private order flow};
    \path [draw, ->] (relay.east) -- (proposer.west);
    \path [draw, ->] (builder.east) -- (-2.9,0);
    \path [draw, ->] (searcher.east) -- (-4.9,0);

    \path [draw, ->] (-6,-0.78) -- (searcher.south);
    \path [draw, ->] (-4,-0.78) -- (builder.south);
    \path [draw, ->] (-6,0.98) -- (-6,0.52);
    \path [draw, ->] (-4,0.98) -- (-4,0.52);
    \path [draw, <-] (-0,0.98) -- (-0,0.32);

    \path [draw, <-] (-2,0.98) -- (-2,0.52);
    \node[shape=circle,draw,inner sep=1pt, font=\footnotesize] (char) at (-5.1,0.2) {1};
    \node[shape=circle,draw,inner sep=1pt, font=\footnotesize] (char) at (-3.1,0.2) {2};
    \node[shape=circle,draw,inner sep=1pt, font=\footnotesize] (char) at (-1,0.2) {3};
    \node[shape=circle,draw,inner sep=1pt, font=\footnotesize] (char) at (-0.2,0.65) {4};
    \node[shape=circle,draw,inner sep=1pt, font=\footnotesize] (char) at (-2.2,0.75) {4};
\end{tikzpicture}

    \caption{PBS scheme visualization. In step \raisebox{.5pt}{\textcircled{\raisebox{-.9pt} {\footnotesize 1}}}, a searcher sends (bundles of) transactions to one or many builders privately, the transactions included in the bundle can be from the public Ethereum mempool, private order flow or from the searcher itself. The builder then builds a high-value block with bundles received from searchers, as well as transactions from the public mempool or private order flow. In step \raisebox{.5pt}{\textcircled{\raisebox{-.9pt} {\footnotesize 2}}}, the builder sends the block to the relay. The relay checks that the block complies with its policies. Then, if requested by the proposer, the relay passes the highest valid bid and corresponding block header on to the proposer in step \raisebox{.5pt}{\textcircled{\raisebox{-.4pt} {\footnotesize 3}}}. The proposer chooses the highest value block amongst the blocks received from the relays, signs the header, and returns it to the relay. This prompts the relay to reveal the full block and broadcast it to the public Ethereum network in step \raisebox{.5pt}{\textcircled{\raisebox{-.4pt} {\footnotesize 4}}}. }
    \label{fig:pbs}
\end{figure}

Figure~\ref{fig:pbs} depicts the PBS scheme and we detail the roles of each of the players in the following: 

\T{Searchers} scan the mempool and look for profitable (bundles of) transactions, generally these transactions involve MEV opportunities, such as sandwich attacks, arbitrage, or liquidations. They pass their bundles to builder(s) privately to avoid them being copied by predatory trading bots scanning the mempool.

\T{Builders} combine bundles received from searchers they are connected to, with private and public transactions to build high-value blocks. Once a block is built they forward the block to one or many relays along with how much they are willing to give the proposer to choose their block -- the bid. In case their block is later chosen by the proposer, the builder's profit is the difference between the block's value and the bid.

\T{Relays} are connected to builders. When they receive blocks from builders, they check that the block is valid, that the payment to the proposer is as indicated in the bid, and that the block complies with any censoring filters implemented by the relay. They identify the block with the highest bid. Then, they send the block header, along with the size of the payment to the proposer. 

\T{Proposers} request blocks from all relays they are connected to. They are expected to choose the block with the highest bid if the value of this bid exceeds the value of the block they built themselves. Once they have chosen a block, they sign the block header received from the relay which will prompt the relay to broadcast the block.\vspace{2pt}

Note that in blocks built through PBS, the block builder is indicated as the fee recipient and the builder transfers the promised amount to the proposer in the block's last transaction. We further highlight that one entity can play all the roles at once. For instance, if a builder also runs a searcher we would refer to this as an \textit{integrated seacher}.

\subsection{Decentralized Exchanges}\label{sec:DEXes}
DEXes are one of the most popular DeFi applications. They allow users to exchange cryptocurrencies on the blockchain without giving up custody of their assets. While traditional CEXes generally utilize a limit order book to match orders, most DEXes are \textit{constant function market makers (CFMM)}. With a CFMM, users execute their order against a \textit{liquidity pool} holding reserves of two or more cryptocurrencies. Trades against the pool execute according to a pre-defined trading function specified in the pool's smart contract, i.e., the pool only accepts trades that keep its trading function constant. Importantly, every trade that executes in a CFMM pool moves the pool's price deterministically, dependent only on the pool's reserves ahead of the trade and the trade's input size. Consequently, if an arbitrage opportunity arises, the assets locked in the liquidity pool determine how large the arbitrage trade can be. The larger the liquidity pools, the larger the required arbitrage trades are to synchronize the prices across different trading venues. 

\subsection{Maximal Extractable Value}\label{sec:MEV}

MEV refers to the profit that can be extracted through including, excluding, and reordering transactions in a block~\cite{DaianFlash2020}. The most commonly observed and measured types of MEV on Ethereum are \textit{sandwich attacks}, \textit{cyclic arbitrage}, and \textit{liquidations}. Succinctly, a sandwich attack has the attacker front- and back-running the victim's transaction on a DEX for a profit. A cyclic arbitrage profits from price differences across DEXes. Liquiditations close under-collateralized loans on lending protocols. The liquidator can buy the collateral of a position at a discount, as a reward for repaying the debt. Oftentimes the liquidator then sells the collateral on a DEX\cite{QinEmpirical2021}.

\section{Data Collection}
To measure the prevalence and impact of non-atomic arbitrage trades on the Ethereum ecosystem, we collect four different types of data. 
Namely, we collect Ethereum blockchain (cf. Section~\ref{sec:EthereumBlockchainData}), PBS relay data (cf. Section~\ref{sec:RelayData}), Ethereum network data (cf. Section~\ref{sec:EthereumNetworkData}), and cryptocurrency price data (cf. Section~\ref{sec:PriceData}). All data we collect from block 15,537,393, i.e., the block of the merge on Ethereum on 15 September 2022, to block 18,473,542, i.e., the last block on 31 October 2023. Thus, our data set covers the entire history of the Ethereum PoS up until 31 October 2023.

\subsection{Ethereum Blockchain}\label{sec:EthereumBlockchainData}
\T{Decentralized Exchange Trades.} We collect data for all DEX trades on Uniswap V2, Uniswap V3, Curve, Balancer and Sushiswap. These five biggest DEXes in terms of total value locked (TVL) on the Ethereum blockchain~\cite{Defillama-ethereum-dexes}. In total, our data set comprises 77,019,583 swaps on DEXes. For each trade, we identify the pool in which the trade was executed as well as the amount of tokens exchanged. We further, record whether the trade was executed successfully, i.e., whether the execution of the transaction was successful. 

\T{Maximal Extractable Value.} To differentiate non-atomic arbitrage trades from types of MEV and to compare their respective prevalence on DEXes, we collect MEV transaction labels from zeromev~\cite{zeromev}. These labels specify whether a transaction was part of a sandwich attack (i.e., victim and attacker transaction), performs a cyclic arbitrage, or a liquidation. 

\T{Rewards Data.} We collect Ethereum execution layer reward data. In particular, for each block we calculate the rewards received by the block's fee recipient: (1) priority gas fees, and (2) tips, i.e., direct transfers to the fee recipient address. Note that the execution layer rewards are also referred to as the block value. We, further, record information regarding the PBS scheme rewards. In the PBS scheme specification, the block builder is indicated as the block's fee recipient and in the block's last transaction transfers the agreed-upon value to the block proposer. Thus, for blocks that follow the PBS specification we record the amount transferred by the fee recipient to the block proposer. 

\subsection{PBS Relays}\label{sec:RelayData}

We collect data from eleven relays: Aestus, Agnostic, Blocknative, bloXroute (Ethical), bloXroute (Max Profit), bloXroute (Regulated), Eden, Flashbots, Manifold, Relayoor and UltraSound. The relays implement public APIs that give access to the blocks delivered by the relays to the proposers, as well as the bids received by the builders. Note that some relay APIs, namely Blocknative, bloXroute (Ethical), Eden, and Relayoor, were not reachable. For these, our data only extends to 31 April 2023, i.e., the last time we scraped the data. Further, bloXroute (Ethical), bloXroute (Max Profit), and bloXroute (Regulated) removed access for builder bids from their APIs. When querying the builder bids API for blocks that passed through the respective bloXroute relay, the response is consistently empty. Thus, for the three bloXroute relays the bid data only extends to 31 January 2023, i.e., the last time we scraped that data and received non-empty responses. 

Importantly, the data by the relays is self-reported and thus requires some trust in the relays, as it cannot be fully verified. However, we combine our Ethereum blockchain reward data set with our PBS data set to verify that the block values reported by the relays correspond to those received by the validators. If this is not the case, we disregard any relay bid data for that block. 

\subsection{Ethereum Network}\label{sec:EthereumNetworkData}
To obtain block times, i.e., the time at which a block was first seen in the network, and to identify private transactions we use Ethereum network data from the Mempool Guru project~\cite{mempool-guru}. The Mempool Guru project runs geographically distributed Ethereum nodes. Each node records the timestamp at which it first saw any block or transaction. The projects data collection method is detailed in~\cite{YangSoK2022}. We take the earliest timestamp across the nodes operated by the Mempool Guru project as the block time and consider a transaction private if not one node saw the transaction before the block it is included in was seen in the network.

\subsection{Cryptocurrency Price}\label{sec:PriceData}
We obtain historical cryptocurrency price data from Binance.com~\cite{binance-data} and CoinMarketCap~\cite{CoinMarketCap2023}. In particular, we obtain candlestick data from Binance.com with an interval of a second as well as aggregate trade data for ETH and BTC. We further obtain daily candlestick data from CoinMarketCap for 60 cryptocurrencies.

\begin{figure*}[t]\vspace{-6pt}
    \centering
    \begin{tikzpicture}[scale=0.98]
  
  \node[draw, rounded corners, minimum width=1.5cm, minimum height=0.7cm,fill=YELLOW, font=\footnotesize] (proposer) at (0, 0)  {proposer};

  \node[draw, rounded corners, minimum width=1.5cm, minimum height=0.7cm,fill=LIGHTCYAN, font=\footnotesize] (relay) at (-4, 0)  {relay};

  \node[draw, rounded corners, minimum width=1.5cm, minimum height=0.7cm,fill=PINK, font=\footnotesize] at (-8, 0) (builder) {builder};
          
  \node[draw, rounded corners, minimum width=1.5cm, minimum height=0.7cm,fill=LIGHTGREEN, font=\footnotesize] at (-12, 0) (searcher) {searcher};
    \path [draw, ->] (relay.east) -- (proposer.west);
    \path [draw, ->] (builder.east) -- (relay.west);
    \path [draw, ->] (searcher.east) -- (builder.west);

    \node[draw, rounded corners, minimum width=5cm, minimum height=1.6cm,fill=lightgrey, font=\footnotesize, text=black,align=left] (step1) at (-12, 1.4)  {};
    \node[minimum width=3cm, minimum height=2cm, font=\footnotesize, text=black,align=left,text width=2.7cm] (step1) at (-11, 1.4)  {\raisebox{.5pt}{\textcircled{\raisebox{-.6pt} {\footnotesize 1}}} a price difference between on-chain DEXes and off-chain markets occurs};
\node[inner sep=0pt] (russell) at (-13.5,1.4)
    {\includegraphics[width =1.7cm]{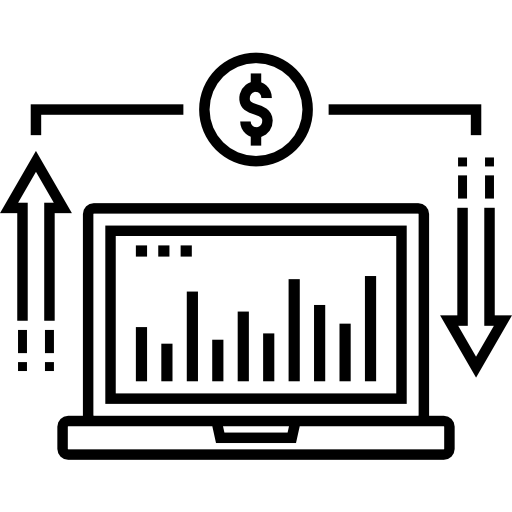}};

    \node[draw, rounded corners, minimum width=2.4cm, minimum height=1.5cm,fill=lightgrey, font=\footnotesize, text=black,align=left] (step2) at (-10,-1)  {};
    \node[minimum width=3cm, minimum height=2cm, font=\footnotesize, text=black,align=left,text width=2.2cm] (step2) at (-10, -1)  {\raisebox{.5pt}{\textcircled{\raisebox{-.6pt} {\footnotesize 2}}} seacher sends non-atomic arbitrage transaction $T_A$ to builder};

    \node[draw, rounded corners, minimum width=2.4cm, minimum height=1.5cm,fill=lightgrey, font=\footnotesize, text=black,align=left] (step2) at (-6,1)  {};
    \node[minimum width=3cm, minimum height=2cm, font=\footnotesize, text=black,align=left,text width=2.2cm] (step3) at (-6, 1)  {\raisebox{.5pt}{\textcircled{\raisebox{-.6pt} {\footnotesize 3}}} builder builds high value block and includes tx $T_A$ from searcher};

    \node[draw, rounded corners, minimum width=2.4cm, minimum height=1.5cm,fill=lightgrey, font=\footnotesize, text=black,align=left] (step2) at (-6,1)  {};
    \node[minimum width=3cm, minimum height=2cm, font=\footnotesize, text=black,align=left,text width=2.2cm] (step3) at (-6, 1)  {\raisebox{.5pt}{\textcircled{\raisebox{-.6pt} {\footnotesize 3}}} builder builds high value block and includes tx $T_A$ from searcher};

    \node[draw, rounded corners, minimum width=2.4cm, minimum height=1.5cm,fill=lightgrey, font=\footnotesize, text=black,align=left] (step2) at (-6,1)  {};
    \node[minimum width=3cm, minimum height=2cm, font=\footnotesize, text=black,align=left,text width=2.2cm] (step3) at (-6, 1)  {\raisebox{.5pt}{\textcircled{\raisebox{-.6pt} {\footnotesize 3}}} builder builds block with $T_A$ and sends it to relay with the bid};

    \node[draw, rounded corners, minimum width=2.4cm, minimum height=1.5cm,fill=lightgrey, font=\footnotesize, text=black,align=left] (step2) at (-2,-1)  {};
    \node[minimum width=3cm, minimum height=2cm, font=\footnotesize, text=black,align=left,text width=2.2cm] (step4) at (-2, -1)  {\raisebox{.5pt}{\textcircled{\raisebox{-.6pt} {\footnotesize 4}}} if requested, relay sends highest value block to proposer};

    \node[draw, rounded corners, minimum width=5cm, minimum height=1.6 cm,fill=lightgrey, font=\footnotesize, text=black,align=left] (step1) at (0, 1.4)  {};
    \node[minimum width=3cm, minimum height=2cm, font=\footnotesize, text=black,align=left,text width=2.7cm] (step5) at (1, 1.4)  {\raisebox{.5pt}{\textcircled{\raisebox{-.6pt} {\footnotesize 5}}} proposer picks highest bidding block with transaction $T_A$ and proposes the block };
\node[inner sep=0pt] (russell) at (-1.5,1.4)   {\includegraphics[width =1.7cm]{Figures/monitor.png}};

\end{tikzpicture}

    \caption{Non-atomic arbitrage illustration. A price difference between DEXes and off-chain markets occurs (step \raisebox{.5pt}{\textcircled{\raisebox{-.6pt} {\footnotesize 1}}}), imagine that the ETH-USDT price is higher on off-chain markets than on-chain markets. In step  \raisebox{.5pt}{\textcircled{\raisebox{-.6pt} {\footnotesize 2}}}, the searcher submits transaction $T_A$ to profit from this price difference to the builder, i.e., buys ETH for USDT on DEXes. The builder then includes transaction $T_A$ (as long as the fees paid by the transaction are sufficient) in the block and forwards the block to the relay along with the bid (step \raisebox{.5pt}{\textcircled{\raisebox{-.6pt} {\footnotesize 3}}}). The relay chooses the block with the highest bid to pass on the to proposer (step \raisebox{.5pt}{\textcircled{\raisebox{-.6pt} {\footnotesize 4}}}). From all relays, the proposer picks the highest value block, and if this block includes  $T_A$ the arbitrage trade executes. Importantly, this arbitrage opportunity is non-atomic as it includes swaps on- and off-chain. }
    \label{fig:nonatomicarb}\vspace{-6pt}
\end{figure*}

\section{Non-Atomic Arbitrage on DEXes Model}

We illustrate the process a non-atomic arbitrageur follows to get the on-chain leg of the trade included in Figure~\ref{fig:nonatomicarb}. Upon detecting an arbitrage opportunity, the searcher submits the desired transaction to the builder. The builder then builds their block and sends it together with their bid to the relay. Next, if requested, the relay sends the highest value block to the proposer who in turn picks the highest bidding block to propose to the network. Importantly, even though visualized separately one entity could for instance operate a builder and a searcher. In this case, we would refer to the latter as an integrated searcher.

\subsection{Non-Atomic Arbitrageur Model}\label{sec:model}

In this section we develop a model for the profit of a non-atomic arbitrageur between two currencies $X$ and $Y$. As discussed in Section~\ref{sec:DEXes}, most DEXes are a CFMM with the largest DEX being Uniswap~\cite{Defillama-ethereum-dexes}. Uniswap currently has two active versions, V2~\cite{adams2020uniswap} and V3~\cite{adams2021uniswap}. Here, we focus on the V2 version, which is also employed by the DEX Sushiswap~\cite{sushiswap-whitepaper}, but the model also holds for V3 between two ticks. The Uniswap CFMM imposes, $x\cdot y = L^2$, where $x$ and $y$ stand for the number of $X$- and $Y$-tokens locked in the liquidity pool, and $L$ is referred to as the liquidity. The (marginal) price $P$ for $X$-tokens in terms of $Y$-tokens is given as $P=y/x$. The reserves can be expressed in terms of the liquidity and price by solving for $x$ and $y$, yielding $x=L/\sqrt{P}$, $y=L \sqrt{P}$.

The price on a DEX on Ethereum is $P_\mathrm{on}$ and the average off-chain price is $\tilde{P}_\mathrm{off}$, i.e., the average price received for an order.\footnote{The average price $\tilde{P}_\mathrm{off}$ can contain transactions on multiple off-chain exchanges. In general, it is the best price attainable for a transaction of a given size.} We denote the price difference between the off-chain and on-chain as $\Delta P = \tilde{P}_\mathrm{off}-P_\mathrm{on}$. Without loss of generality, we assume the searcher discovers $\Delta P > 0$, i.e., the price of $X$-tokens off-chain is higher such that the arbitrageur buys $X$-tokens on the DEX. During the trade, they pay $\Delta y$ of which $f \cdot \Delta y$ is paid in fees to the liquidity providers, where $f$ is the protocol fee. With this and the constant product $L^2=x y$, the amount of $X$-tokens the trader receives is given by
$$\Delta x = \tfrac{x(1-f) \Delta y}{y+(1-f)\Delta y}.$$
Apart from $f$, the arbitrageur is charged a fee by the off-chain exchanges which we denote as $g$. 

As long as the off-chain price remains higher than the on-chain price, the arbitrageur can extract more value by buying more $\Delta x$ on the DEX. Thus, the arbitrageur will trade on the DEX until the price they receive for an infinitesimal unit of $X$ is $\tilde{P}_\mathrm{off}  (1-g)$.\footnote{Here we assume that the off-chain price is unaffected by the arbitrage. In practice, the arbitrageur would receive a quote for a given trade size and can be used as $\tilde{P}_\mathrm{off}$.} The pool's marginal price at the end of the trade is then given by
$$P_\mathrm{end} =\tilde{P}_\mathrm{off}  (1-g)  (1-f)= \tfrac{y+(1-f)\Delta y}{x-\Delta x}.$$
Simultaneously, to this buy order, the arbitrageur sells the same amount of $X$-tokens, $\Delta x$ through the off-chain exchanges at an average price of $\tilde{P}_\mathrm{off}$.

\begin{figure}[h]
    \centering
    \includegraphics[scale=1]{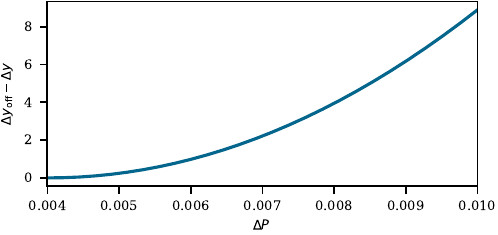}
    \caption{Arbitrageur profit as a function of the price difference for $L =10^6$, $P_\mathrm{on}=1$, $f=0.3\%$, and $g=0.1\%$. Note we only plot the profit in the profitable range. }
    \label{fig:theory}\vspace{-4pt}
\end{figure}

Thus, the arbitrageur receives 
$$\Delta y_\mathrm{off}=\tilde{P}_\mathrm{off}(1-g) \Delta x =   \tfrac{x \Delta y}{x-\Delta x} ,$$
where we used the previous two equations. Finally, we can compute the arbitrageurs profit by solving $\Delta P$ for $\Delta y$ and inserting it in the above equation for $\Delta y_\mathrm{off}$ which leads to
$$\Delta y_\mathrm{off}-\Delta y = L \tfrac{\left(\sqrt{P_\mathrm{on}}- \sqrt{(1-f)(1-g) (\Delta P + P_\mathrm{on})}\right)^2}{(1-f) \sqrt{P_\mathrm{on}}} $$
where we also expressed $x$ and $y$ in terms of $L$ and the starting on-chain price $P_\mathrm{on}$. The profit as a function of the price change $\Delta P$ is plotted in Figure~\ref{fig:theory}. The profit becomes significant for larger price changes, showing that sufficient price difference will attract non-atomic arbitrage trades. Note that profitable arbitrage only exists for $P_\mathrm{on} <\tilde{P}_\mathrm{off} (1-f) (1-g)$, as otherwise the fees paid to the liquidity pool exceed the price difference that can be profited from. 

We further note that while an arbitrageur still needs to pay the builder to have their trade included, the result shows how valuable the arbitrage trade is for the searcher and limits the bid they would be willing to submit.

\subsection{Case study of Block 18,360,789}

\begin{figure}[t]
    \centering
    \includegraphics[scale=1]{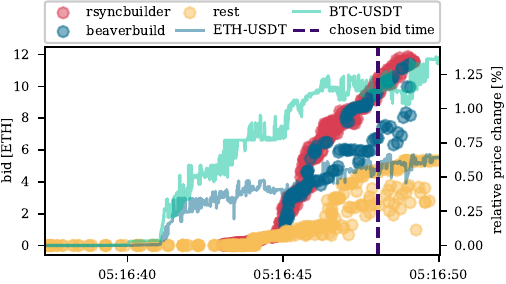}
    \caption{Bids submitted by builders for block 18,360,789 on 16 October 2023, along with the ETH-USDT and BTC-USDT price on Binance.com. Notice that the bids, especially those previously identified ``HFT'' builders~\cite{Gupta2023Centralizing} start to rise significantly shortly after the prices on Binance.com start to move. Importantly, this price movement creates an arbitrage opportunity between DEXes on Ethereum and Binance.com. Further, the bids submitted by HFT builders exceed those by the remaining builders significantly. }
    \label{fig:bidexample}\vspace{-6pt}
\end{figure}

To provide a better understanding of non-atomic arbitrage, we go through a case study of block 18,360,789 -- a block with a significant price change in the lead-up to the block, i.e., the time between the previous block proposal and the block proposal itself. In Figure~\ref{fig:bidexample}, we plot the time and the value of bids from the builders. Note that we hightlight bids from rsyncbuilder (shown in red) and beaverbuild (shown in blue), as these two builders were previously identified as having integrated searchers that perform these non-atomic arbitrage trades~\cite{Gupta2023Centralizing}. The bids from all remaining builders are shown in yellow. We further indicate the time of the bid corresponding to the block that was chosen by the proposer, i.e., the relays deliver the highest bid to the proposer when the proposer requests it. Finally, we also plot the relative price change of ETH-USDT and BTC-USDT on Binance.com during the same time. Recall that USDT is a stablecoin pegged to the \$.

One immediately notices that the time in the lead-up to the block is characterized by extremely high price volatility of both ETH and BTC in comparison to USDT. The relative price change exceeds 0.6\% and 1.3\% in less than ten seconds respectively. This price change creates an arbitrage opportunity, as the prices on DEXes do not move in between blocks. Thus, as outlined in the previous section, the first transaction in a DEX pool can close this price difference and thereby profit from the arbitrage opportunity.  

\begin{table}[t]
\scriptsize

\centering

\begin{adjustbox}{width=\linewidth}
\begin{tabular}{@{}rllllr@{}}
\toprule
tx index & searcher & pool & tokenIn & tokenOut & amount [\$] \\
\midrule
0 & rsyncsearcher3 & \texttt{0x88e6a...f5640} & USDC & ETH & 2148195.308 \\
1 & rsyncsearcher3 & \texttt{0x99ac8...abc35} & USDC & BTC & 659507.701 \\
2 & rsyncsearcher3 & \texttt{0x4585f...a20c0} & ETH & BTC & 554524.411 \\
3 & rsyncsearcher3 & \texttt{0x9db9e...8425b} & USDT & BTC & 186858.044 \\
4 & rsyncsearcher3 & \texttt{0x11b81...697f6} & USDT & ETH & 523225.597 \\
5 & rsyncsearcher3 & \texttt{0x4e68c...dfa36} & USDT & ETH & 524150.487 \\
6 & rsyncsearcher3 & \texttt{0xcbcdf...d62ed} & ETH & BTC & 495617.185 \\
8 & searcher1 & \texttt{0x8ad59...6e6d8} & USDC & ETH & 562787.652 \\
9 & rsyncsearcher3 & \texttt{0x60594...5a270} & DAI & ETH & 116090.952 \\
10 & rsyncsearcher3 & \texttt{0xc2e9f...a25f8} & DAI & ETH & 105871.055 \\
12 & searcher1 & \texttt{0xd51a4...aae46} & USDT & BTC & 17463.007 \\
14 & rsyncsearcher3 & \texttt{0x391e8...86649} & DAI & BTC & 9214.814 \\
17 & rsyncsearcher3 & \texttt{0x06da0...84553} & USDT & ETH & 18283.993 \\
18 & rsyncsearcher3 & \texttt{0xc3d03...5882f} & DAI & ETH & 10533.977 \\
19 & rsyncsearcher3 & \texttt{0x397ff...7aca0} & USDC & ETH & 14457.825 \\
21 & rsyncsearcher3 & \texttt{0xceff5...d3a58} & ETH & BTC & 15450.829 \\
22 & rsyncsearcher3 & \texttt{0x824a3...c0673} & ETH & BLUR & 11451.829 \\
23 & rsyncsearcher3 & \texttt{0x56534...a83b2} & USDT & BTC & 4822.975 \\
27 & rsyncsearcher3 & \texttt{0xcbfb0...7ef3c} & USDC & BTC & 1538.249 \\
28 & rsyncsearcher3 & \texttt{0x290a6...f5b42} & ETH & MATIC & 14629.722 \\
31 & searcher5 & \texttt{0xcd828...a9fbf} & FXS & ETH & 6247.832 \\
33 & rsyncsearcher3 & \texttt{0x919fa...daf79} & ETH & CRV & 4954.943 \\
35 & rsyncsearcher3 & \texttt{0x34704...08d60} & USDT & UNI & 1106.016 \\
36 & rsyncsearcher3 & \texttt{0xa3f55...dfd74} & LDO & ETH & 9727.578 \\
43 & rsyncsearcher3 & \texttt{0x9febc...c82ae} & USDC & 1INCH & 516.780 \\
44 & rsyncsearcher3 & \texttt{0x1d420...bd801} & ETH & UNI & 6336.821 \\
\bottomrule
\end{tabular}

\end{adjustbox}

    \caption{Non-atomic arbitrages identified in block 18,360,789 on 16 October 2023. There are 26 non-atomic arbitrage swaps with a total volume of \$6,023,565 executed by merely three searchers (cf.~\cite{anonymous2023} for a mapping from address to searcher name). Further, rsyncbuilder's integrated searcher is responsible for 23 of these arbitrages. We indicate the tx index, the searcher, pool, and the tokenIn (the token that was sold), the tokenOut (the token that was bought), and the amount (the trade volume in \$). 
    }
    \label{tab:block}\vspace{-6pt}
\end{table}

Coming back to Figure~\ref{fig:bidexample}, we can observe that around five seconds after the price starts to change on Binance.com the bids start to increase. At this point, the price difference appears to be big enough for non-atomic arbitrage to be profitable. Further, we find that bids from the builders that are associated with non-atomic arbitrage transactions are higher than those from the rest and that the bids continue to increase as the ETH and BTC prices on Binance.com increase.  The block chosen by the proposer was built by rsyncbuilder and its bid, i.e., the value received by the proposer was a staggering 10.32~ETH ($\approx$\$23,000 at the time of this writing). In Table~\ref{tab:block}, we take an in-depth look a the non-atomic arbitrage trades we identified with our heuristics that we will introduce in the following Section~\ref{sec:heuristic}. In the block, there were 26 transactions with volume exceeding \$6~million that we identified as performing non-atomic arbitrage trades. Remarkably, all but three of these transactions were by the rsyncsearcher3 -- a searcher we identified to be linked to the rsyncbuilder that won the block (cf. Section~\ref{sec:verticalIntegration}). Additionally, we highlight that the rsyncsearchers paid a remarkable 11.15~ETH (i.e., more than the PBS bid) for their transactions. Thus, the rsyncbuilder built a high-value block largely attributed to fees paid by trades we labeled as non-atomic arbitrage by its own integrated searcher.

Note that most of trades buy ETH or BTC for a stablecoin. As we saw in Figure~\ref{fig:bidexample} the prices of these two cryptocurrencies rose on Binance.com, and they were thus trading for cheaper on DEXes at the beginning of the block. For example, if we look at the price at the beginning of the block, we see that the price of the ETH-USDT Uniswap V3 pool was 1563.57, and at the end of the block, it was 1574.15. The Binance.com price was 1564.61 at the beginning of the block's slot, while it rose to 1574.63 by the time the block was proposed. Notice that the arbitrageurs drove the price almost exactly to the off-chain price, as we show to be optimal in our previous analysis (cf. Section~\ref{sec:model}). Additionally, some transactions exchange cryptocurrencies for cryptocurrencies (i.e., no stablecoins), such as the transaction at index 2 and 6, which exchange ETH for BTC. The relative price increase measured in USDT of BTC was higher than that of ETH. Thus, BTC relative to ETH was also available for cheaper on DEXes at the beginning of the block. Hence, the searcher also swaps ETH for BTC.

\section{Identifying Non-Atomic Arbitrage Trades}\label{sec:heuristic}

With our acquired understanding of non-atomic arbitrage, we now describe our procedure to identify non-atomic arbitrage trades on the Ethereum blockchain. We start by noting that identifying this type of MEV transaction is more difficult and less clear-cut than other types of MEV such as sandwich attacks, cyclic arbitrage or liquidations previously measured on Ethereum blockchain~\cite{QinQuantifying2022,ZhouHigh2021,FerreiraFrontrunner2021,WangCyclic2022,QinEmpirical2021}. All transactions related to these types of MEV can be found on the same blockchain. Importantly, we only observe one side of the arbitrage on the Ethereum blockchain and have practically no visibility into the other side of the arbitrage. Presumably, as mentioned previously the other side of the arbitrage executes on a CEX, as these have the most significant liquidity for cryptocurrency trading. Thus, we do not have access to de-anonymized data and the detail of data available also differs highly between CEX. Additionally, it is also possible that the other side of the arbitrage executes on a different blockchain, but here we are also not able to link the identity of wallets on different chains. 

However, we are still able to infer whether a transaction is likely to be a non-atomic arbitrage. The main assumption that guides are following heuristics is as follows: 

\T{Assumption.} Non-atomic arbitrage transactions have a high value, i.e., a significant profit can be made. Thus, searchers performing these non-atomic arbitrages are likely to protect their transaction (i.e., submit it privately such that it cannot be copied) and pay suspicious amounts to have their ``simple'' swap included (i.e., tip the miner through significant priority fees or direct transfers).\vspace{2pt}

Guided by this assumption we apply the following five heuristics to identify non-atomic arbitrage transactions. 

\T{Heuristic 1.} The transaction is a simple swap, i.e., (1) executes exactly one swap in a DEX, (2) is not labeled as a sandwich attack (frontrun, backrun, or victim), a DEX arbitrage, or a liquidation, and (3) does not use more than 400,000 in gas.

\T{Heuristic 2.} The transaction is private, i.e., it did not enter the mempool before the block was propagated. 

\T{Heuristic 3.} The transaction includes a coinbase transfer to the fee recipient or its priority fee is at least 1 GWei. 

\T{Heuristic 4.} The swap executed by the transaction is either the first swap that executes in the pool in its direction, or all preceding transactions had the same recipient. 

\T{Heuristic 5.} The swap exchanges two established tokens, i.e., traded on CEXes (cf.~\cite{anonymous2023} for a list). \vspace{1pt}

Heuristic 1 ensures that the transaction only executes a ``simple'' swap on a DEX and that we do not mix it up with another type of high-value MEV transaction.\footnote{Note that we limit the size of the transaction in units of gas to ensure that the transaction does not execute further complicated logic we have no insights into. Without the gas limit we picked up transactions rebalancing liquidity positions on Uniswap V3. The limit excludes these transactions but is large enough for ordinary swaps on all five DEXes. A simple swap on in Curve V2 pools consumes the highest amount of gas for the DEXes we tracked at $\approx$350,000 gas, whereas rebalancing a liquidity position on Uniswap V3 uses $\approx$450,000 gas.} Together, heuristics 2 and 3 allow us to identify that the transaction is valuable as it was submitted privately and tips the validator.\footnote{For heuristic 3, we do not impose a minimum on the value of the coinbase transfer as less than 5\% of transaction make such a transfer. Additionally, we set the minimum priority fee limit to 1 GWei (i.e., approximately the median priority fee of DEX swaps) to ensure that the transaction is paying a significant tip. We do not set the limit higher as beaverseachers (previously identified non-atomic arbitrage searchers) use a priority fee of 1 Gwei for more than 50\% of their transactions without coinbase transfers.} With heuristic 4 we ensure that the transaction is the first transaction to execute in the respective direction in the pool. If this were not to be the case, the price in the pool would already have moved unfavorably for the transaction, and the sought-after arbitrage transaction might already have been closed. Finally, heuristic 5 ensures that there is likely to be another market for the trade executed, i.e., that it is possible to make the other side of the trade on a CEX for instance. Note that a transaction that violates heuristics 2, 3, 4, or 5 might still be a non-atomic arbitrage, but in that case, is likely to be a less sought-after opportunity. However, wherever possible we are conservative with what we label as non-atomic arbitrage. We also note that a transaction that does not perform any arbitrage could fulfill all our heuristics. 

\begin{table*}[t]\vspace{-6pt}
\scriptsize

\centering

\begin{adjustbox}{width=\linewidth}
\begin{tabular}{@{}lrcccccccc@{}}
\toprule
 &  total swaps & simple   & private &first swap in pool  & top tokens  & coinbase transfer & priority fee &coinbase transfer or priority fee  & all \\
& & heuristic 1 & heuristic 2 & heuristic 4&heuristic 5 & & &heuristic 3\\
\midrule
beaversearchers & 1,188,696 & 0.982 & 0.815 & 0.826 & 0.822 & 0.608 & 0.313 & 0.917 & 0.580 \\
rsyncsearchers & 346,570 & 1.000 & 0.997 & 0.924 & 0.931 & 0.059 & 0.895 & 0.954 & 0.884 \\
mantasearcher & 60,813 & 0.999 & 0.941 & 0.921 & 0.955 & 0.940 & 0.057 & 0.977 & 0.879 \\
overall & 77,019,583 & 0.651 & 0.317 & 0.280 & 0.151 & 0.036 & 0.568 & 0.601 & 0.028 \\
\bottomrule
\end{tabular}
\end{adjustbox}

    \caption{Proportion of transactions from known integrated searcher (e.g., searchers associated with beaverbuild, mantabuilder, and rsyncbuilder) our five heuristics apply to individually and together in comparison to all swaps on DEXes. }
    \label{tab:heuristic}\vspace{-6pt}
\end{table*} 

Thus, to test our heuristics we measure the proportion of transactions our heuristic applies to for integrated searchers previously identified to be performing non-atomic arbitrage on the Ethereum blockchain (e.g., searchers associated with beaverbuild, mantabuilder and rsyncbuilder)~\cite{Gupta2023Centralizing} and compare that figure to all transactions in our data collection period. In Table~\ref{tab:heuristic}, we record what proportion of transactions each of our heuristics applies to. The last column in Table~\ref{tab:heuristic} indicates what proportion of transactions from these integrated searchers as well as overall our heuristics capture. We find that our heuristics apply to 58.0\%, 87.9\%, and 88.4\% of transactions from the integrated searchers associated with beaverbuild, mantabuild, and rsyncbuilder respectively, but only apply to 2.8\% of transactions overall. 

When considering the integrated searchers from the three builders known to perform non-atomic arbitrage, we notice that our heuristics apply to nearly 90\% of transactions of mantasearcher and the rsyncsearchers, but only around 60\% of transactions from the beaversearchers. Interestingly, mantasearcher tips through coinbase transfers whereas rsyncbuild tips through priority fees. Transactions from beaversearchers are especially less likely to be private (heuristic 2), the first swap that executes in a pool (heuristic 4), and less likely to swap two established tokens (heuristic 5). When analyzing transactions by beaverbuild that do not meet the heuristic, we find that sometimes it appears that they still perform the non-atomic arbitrages, but that these are less valuable, i.e., swap less known tokens and/or only transfer smaller volumes. Thus, in these cases, it might also be less important for beaverbuild to protect their transaction, i.e., do not send the transaction privately. In other cases, it appears that beaverbuild is not performing non-atomic arbitrages, failed at performing the non-atomic arbitrage or again only performs a less valuable non-atomic arbitrage, as at least one swap that moves the price in an unfavorable direction for beaverbuilds swap precedes it. Finally, we find that for one month, the beaversearchers were not paying tips, but were instead subsidized by the beaverbuilder (cf. Appendix~\ref{app:beaverbuild}). Thus, exclusively for beaversearchers, we do not require them to fulfill heuristic 3, if the swap executes in a block built by the beaverbuilder. In doing so, 60.4\% of swaps by the beaversearchers fulfill these tailored heuristics. We note that we likely do not capture all non-atomic arbitrage transactions from beaversearchers as our heuristics are potentially too strict, but stick to our heuristic to avoid overstating the prevalence of this type of arbitrage.

Finally, we note again that our heuristics apply to the (vast) majority of transactions from the previously identified seachers, but only to 2.8\% of swaps overall. Further, half of these transactions are swaps that meet all heuristics from the searchers associated with beaverbuild, mantabuilder, and rsyncbuilder. Thus, we believe that our heuristics are well-tailored to identifying non-atomic arbitrage. We will further show in the following that eleven searchers are responsible for more than 80\% of these identified non-atomic arbitrage trades, an additional indicator that our heuristics are well-suited to capture non-atomic arbitrage.

\section{Analyzing Non-Atomic Arbitrage}
In the following, we analyze the non-atomic arbitrage identified with our previously introduced heuristics. 
\subsection{Non-Atomic Arbitrage Trade Landscape}

We start by providing a general overview of the non-atomic arbitrage on DEXes. On average, we identify 5216 trades as non-atomic arbitrages per day during our data collection window that spans a little over a year. This figure is in comparison to 5219 sandwich attacks (front- and back-running transaction counted as one), 1245 cyclic arbitrage (i.e., atomic on-chain arbitrage), and 14 liquidations. Recall, that sandwich attacks, cyclic arbitrage, and liquidations are considered the most common and well-studied types of MEV. Thus, non-atomic arbitrage, in terms of the number of occurrences is the second most prevalent type of MEV out of these four on the Ethereum blockchain. We provide a more detailed comparison between non-atomic arbitrage and these other types of MEV in Appendix~\ref{app:mev}.

\begin{figure}
    \centering
    \includegraphics[scale=1]{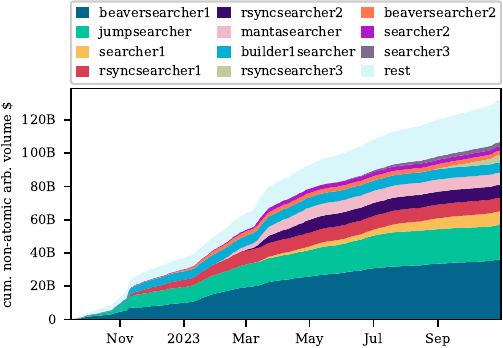}
    \caption{Cumulative volume of non-atomic arbitrage swaps by searcher. In total, the non-atomic arbitrage volume during our data collection period is \$132~billion. We provide a mapping from the searcher address to the searcher name as part of our data set~\cite{anonymous2023}.}
    \label{fig:cumulativeVolume}\vspace{-6pt}
\end{figure}

In Figure~\ref{fig:cumulativeVolume}, we consider the cumulative volume of non-atomic arbitrage over time and highlight the eleven biggest searchers performing non-atomic arbitrage by volume. We provide a mapping from searcher name to address in our data set~\cite{anonymous2023}. Further note that some searchers (beaversearcher1, beaverseacher2, rsyncseacher1, rsyncseacher2, rsyncseacher3, and mantaseacher) have names that suggest that they are linked to large builders in the PBS scheme (beaverbuild, rsyncbuilder and mantabuilder)~\cite{HeimbachEthereum2023}. This naming is due to our suspicion that the respective builders and searchers are controlled by the same entities and we will detail our reasoning in Section~\ref{sec:verticalIntegration}. 

Returning to Figure~\ref{fig:cumulativeVolume}, in total, the volume of trades we identify as non-atomic arbitrage are responsible for \$132~billion on DEXes, e.g., Uniswap V2, Uniswap V3, Sushiswap, Curve, and Balancer. Startlingly, the total volume on these five DEXes in the same period of time is \$460~billion~\cite{Defillama-api}. Thus, the identified non-atomic arbitrage accounts for nearly 30\% of the total volume of the five biggest DEXes on the Ethereum blockchain. 

We reiterate here that there is no clear-cut way of identifying non-atomic arbitrage, as we lack data transparency regarding the off-chain. However, the fact that merely eleven searchers are responsible for 80\% of the non-atomic arbitrage volume identified in our analysis and that we identify the majority of their transaction as non-atomic arbitrage trades for all but one searcher (cf. Appendix~\ref{app:heuristics}), leads us to believe that our heuristics are well-suited to identify non-atomic arbitrage transactions. Additionally, these eleven searchers can further be clustered into eight entities, three searchers are associated with rsyncbuilder and two searchers are associated with beaverbuild (cf. Section~\ref{sec:verticalIntegration} for a detailed description of the searcher/builder relationship identification). Thus, a very small number of entities is responsible for the vast majority of the non-atomic arbitrage volume on DEXes. In fact, the two biggest searchers, i.e., beaversearcher1 and jumpsearcher, account for 49.2\% non-atomic arbitrage volume. We further point out that many of these large searchers (beaversearchers, jumpsearcher, rsyncsearchers, and mantasearcher) are rumored to be linked to HFT firms operating in traditional finance~\cite{Gupta2023Centralizing,2022jump}, a further indicator that our heuristics capture non-atomic arbitrage as well as a sign that big players from traditional finance turning to DeFi for profits.

Finally, we highlight that two periods of time exhibit significant increases in the cumulative volume: (1) mid-November 2022, which is a period of time characterized by high volatility of cryptocurrencies during the FTX bankruptcy~\cite{ftx-collapse}, and mid-March 2023, which corresponds to the USDC depeg and high cryptocurrency price volatility~\cite{usdc-depeg}. High cryptocurrency price volatility --- leading to price differences between cryptocurrency markets --- drives non-atomic arbitrage opportunities and these dramatic increases in non-atomic arbitrage volume during periods of high volatility are thus expected. We provide a more in-depth analysis of the correlation between price volatility and non-atomic arbitrage volume in Section~\ref{sec:drivers}.

\begin{figure}
    \centering
    \includegraphics[scale=1]{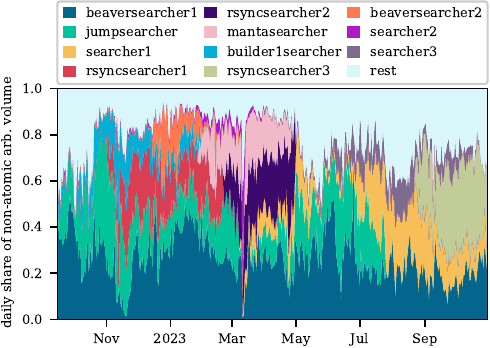}
    \caption{Daily share of the non-atomic arbitrage volume accounted for by the biggest eleven searchers. Notice that large searchers go in and out of operation over time and that on most days two searchers are responsible for more than half the volume.}
    \label{fig:shareOfArbitrages}\vspace{-6pt}
\end{figure}

In regards to Figure~\ref{fig:cumulativeVolume}, we conclude that the non-atomic arbitrage volume on DEXes is immense and controlled by a few large entities. To gain a better understanding of the evolution over time, we take a more in-depth look at the share of non-atomic arbitrage volume controlled by the large searchers in Figure~\ref{fig:shareOfArbitrages}. We start by noting that on nearly 75\% of days, merely two searchers are responsible for at least 50\% of the arbitrage volume -- underlying the high concentration in the non-atomic arbitrage market. Interestingly, beaverseacher1, the biggest non-atomic arbitrage searcher, is the only major searcher operating through our entire data collection period. Generally, beaversearcher1 accounts for at least 10\% of the volume except for during the two time periods characterized by very high cryptocurrency price volatility in the previous, i.e., mid-November 2022 and mid-March 2023. Potentially, the non-atomic arbitrage market becomes more competitive during days characterized by exceptional events in the blockchain ecosystem, and beaverseacher1 loses some market share as a result. The second biggest searcher, jumpsearcher, on the other hand, was operating since the merge but stopped in late July 2023, while searcher1, the third biggest searcher, only started operations in mid-March 2023 and from then on was operating throughout our entire data collection window. Interestingly, the three searchers associated with rsyncbuilder all operate during non-overlapping time windows, it appears that one searcher is replacing the other -- potentially upgrades to the searcher smart contract. Furthermore, there was a period of time between May and September 2023, when no searcher we could associate with rsyncbuilder was operating.  

\begin{table*}[t]
\scriptsize

\centering

\begin{adjustbox}{width=\linewidth}
\begin{tabular}{@{}lrrrrrrrrrrr@{}}
\toprule
 & beaversearcher1 & jumpsearcher & searcher1 & rsyncsearcher1 & rsyncsearcher2 & mantasearcher & builder1searcher & rsyncsearcher3 & beaversearcher2 & searcher2 & searcher3 \\
\midrule
proportion of trades & 0.704 & 0.865 & 0.867 & 0.867 & 0.721 & 0.814 & 0.661 & 0.673 & 0.634 & 0.893 & 1.000 \\
proportion of volume & 0.916 & 0.901 & 0.933 & 0.922 & 0.898 & 0.937 & 0.813 & 0.855 & 0.912 & 0.973 & 1.000 \\
\bottomrule
\end{tabular}
\end{adjustbox}

    \caption{Proportion of trades and value for each of the top searchers in pools between ETH, BTC, USDC, USDT, and DAI.}
    \label{tab:topTokens}\vspace{-6pt}
\end{table*} 

\begin{figure*}[t]
    \centering
    \includegraphics[scale=1]{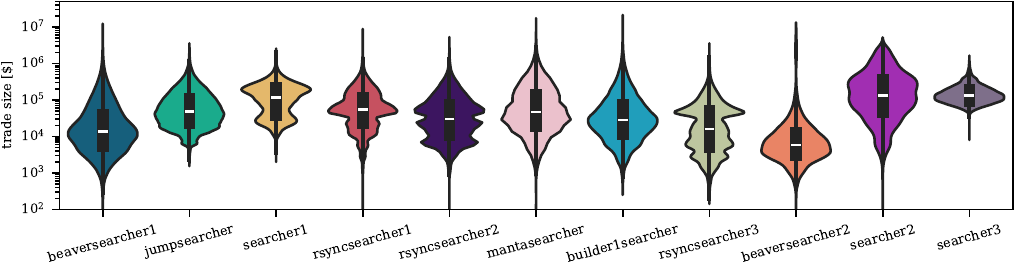}
    \caption{Distribution of the trade size of non-atomic arbitrage transactions of each of the eleven biggest searchers in terms of volume. The average trade size across these eleven searchers is \$73,891. }
    \label{fig:tradeSizebyBot}\vspace{-6pt}
\end{figure*}

To get a better understanding of the differences in strategies utilized by the various searchers, we plot the distribution of trade sizes for each searcher as a violin plot in Figure~\ref{fig:tradeSizebyBot}. A violin plot includes a boxplot, which indicates the lower quantile, median, and upper quantile, as well as a kernel density plot to visualize the distribution of values. We observe clear differences in the trade size distribution between searchers (cf. Figure~\ref{fig:tradeSizebyBot}). For instance, beaversearcher1's trade size distribution is quite smooth with the peak and median being around \$10,000 and trade sizes ranging from less than \$100 to more than \$10~million. Interestingly, except beaverseacher2, all other major searchers have larger median trade sizes for their non-atomic arbitrage swaps. We further note that while their median trade sizes are slightly higher than that of beaverbuild, mantaseacher and builder1searcher exhibit similar trade size distributions to beaversearcher1.  A couple of searchers (e.g., jumpsearcher, searcher1, searcher2, and searcher 3) appear to be specialized in high-volume non-atomic arbitrage trades. The median trade sizes of the former three all exceed \$100,000. Finally, we observe that the three rsyncbuilders tend to have several peaks in their trade size distributions, i..e, they seem to be biased towards several distinct trade sizes. Further, we observe that with each iteration of the rsyncsearcher, the median trade size declines. While we cannot be sure what the cause of this is, it could be a sign that with time rsyncsearcher is turning towards smaller and likely less fought-after (as they are less valuable) non-atomic arbitrage transactions. 

In Table~\ref{tab:topTokens}, we further analyze what proportion of non-atomic arbitrage volume and swaps by each of the largest eleven searchers are in pools between ETH, BTC, USDC, USDT, and DAI. These are the biggest cryptocurrencies on DEXes in terms of the number of swaps. One would assume that a significant proportion of volume is in such pools, given that these cryptocurrencies also have liquid markets on both DEXes and CEXes.  We find that for all searchers but searcher3, who has exclusively had non-atomic arbitrage trades between these five tokens, the proportion of trades between these five tokens is smaller than the respective proportion of volume (cf. Table~\ref{tab:topTokens}).  This discrepancy indicates that higher volume non-atomic arbitrage trades are executed between these five tokens, which is expected given that comparatively the DEX pools for these five tokens have higher liquidity. Therefore, it requires greater trade sizes to move the price as outlined in Section~\ref{sec:model}. Except for searcher3, we observe largely similar behavior in terms of the proportion of volume/trades executed between these five highly liquid tokens or otherwise for all searchers. Interestingly, we notice again that with each iteration of the rsyncsearchers, the proportion of trades/volume executed between the top five tokens drops. Thus, it again appears as if with time the rsyncbuilder is turning to potentially less fought-after non-atomic arbitrage transactions.

\subsection{Drivers of Non-Atomic Trades}\label{sec:drivers}

\begin{figure}
    \centering
    \includegraphics[scale=1]{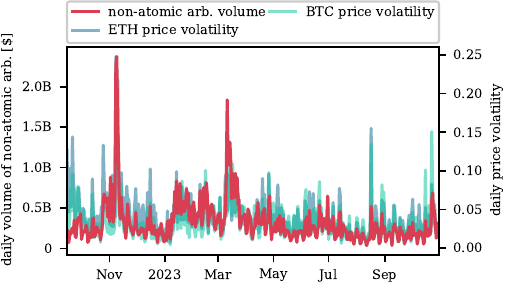}
    \caption{Daily volume of non-atomic arbitrage trades along with the daily volatility of the ETH and BTC price. The correlation between the non-atomic arbitrage volume and the ETH and BTC volatility is 0.725 and 0.729 respectively.}
    \label{fig:numberHFTTrades}\vspace{-6pt}
\end{figure}

In the following, we investigate the drivers of non-atomic arbitrage trades on DEXes. Recall, that non-atomic arbitrage leverages price differences between two markets (likely an on-chain DEX and an off-chain CEX), and these are expected to be heightened in times of high cryptocurrency price volatility. In more detail, price volatility on CEXes can drive non-atomic arbitrage transactions, as it signals a divergence from the price at an on-chain DEX. Thus, we commence by investigating the relationship between the daily volume of trades we identified as non-atomic arbitrage trades and the daily price volatility of BTC and ETH in Figure~\ref{fig:numberHFTTrades}. Note that we measure the BTC and ETH price volatility since the change in cryptocurrency prices is closely correlated to changes in the two~\cite{Blockworks-crypto-correlation}. 
We measure the daily volatility by 
$\log_{10} \left({P_{\text{high}}}/{P_{\text{low}}}\right),$
where $P_{\text{high}}$ and $P_{\text{low}}$ are the daily high and low of the price respectively.

Looking at Figure~\ref{fig:numberHFTTrades}, we observe that days with a high volume of non-atomic arbitrage trades tend to correspond to days with high BTC and ETH price volatility. Additionally, the highest volume days correspond to two periods identified earlier: the FTX collapse in November 2022 and the USDC depeg in March 2023. We further note that the correlation between the daily ETH price volatility and the volume of non-atomic arbitrage is 0.725 with a p-value of $1.30\cdot 10^{-69}$ -- inferring extremely high statistical significance of the correlation. For the BTC price volatility, the correlation is even higher at 0.729 with a p-value of $2.86\cdot 10^{-68}$.  

Note that the strong correlation between price volatility and non-atomic arbitrage volume does not necessarily imply causation. To further explore relationship, we calculate the correlation between volume (excluding non-atomic arbitrage) and non-atomic arbitrage volume. The correlation is 0.655, and thus, lower than the correlation between non-atomic arbitrage volume and price volatility of 0.725 and 0.729 for ETH and BTC respectively. The stronger correlation between non-atomic arbitrage volume and price volatility, underlines the relationship between the two. Additionally, we note that the correlation is lower despite our heuristics being conservative and likely undercounting the prevalence of non-atomic arbitrage, i.e., some remaining volume is probably non-atomic arbitrage and thereby correlated. Thus, price volatility of ETH and BTC on CEXes likely drives non-atomic arbitrage on DEXes.

\begin{figure}
    \centering
    \includegraphics[scale=1]{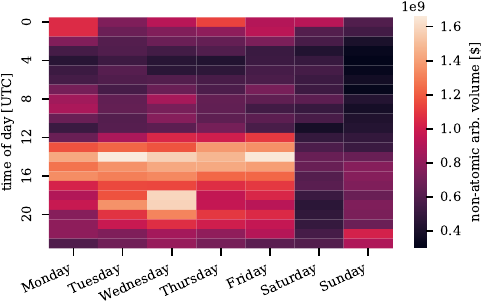}
    \caption{Non-atomic arbitrage volume by weekday and time of day (UTC). Times of increased volume correspond to US, Asian, and European market hours.}
    \label{fig:timeOfDay}\vspace{-6pt}
\end{figure}

In Figure~\ref{fig:timeOfDay}, we visualize the volume of non-atomic arbitrage trades by weekday and hour using UTC as the timezone. Observe that the highest volume in non-atomic arbitrage trades is seen Monday through Friday between 14:00 UTC and 21:00 UTC, which corresponds to US market openings. Especially US market opening at 14:30 UTC is associated with increased non-atomic arbitrage volume. An additional, less noticeable cluster of non-atomic arbitrage volume is observed Monday through Friday around 0:00 UTC, which corresponds to the opening of the Asian markets, and even less noticeable Monday through Friday at 8 am UTC -- European market opening. This heightened volume during market hours and especially during the respective market openings is expected, as cryptocurrency prices are correlated with the US markets \cite{bitcoinCorrelation}. Thus, volatility in cryptocurrency prices is expected to increase during (US) market trading hours. Hence, we see a rise in the volume of non-atomic arbitrage trades during these times, as cryptocurrency prices are most volatile. We observe a final and interesting peak in non-atomic arbitrage volume on Wednesdays around 18:00 UTC. This corresponds to the time of the press conference following the  Federal Open Market Committee Meeting at 19:00 UTC and the preceding release of notes at 18:30 UTC~\cite{2022fomc}. In this meeting decisions regarding the federal interest rate are made and they have led to significant market and cryptocurrency price movement~\cite{2022fed}.

To conclude, we identified the expected link between cryptocurrency price volatility and non-atomic arbitrage. 

\subsection{Integrated Seachers}\label{sec:verticalIntegration}

As mentioned previously, the searchers looking for and executing non-atomic arbitrage opportunities and the builders in the PBS scheme building the block need not be separate entities. In fact, previous work~\cite{Gupta2023Centralizing} has demonstrated that three builders (i.e., beaverbuild, rsyncbuilder, and mantabuilder) are likely to operate their own non-atomic arbitrage searchers but did not establish links between these builders and specific searchers. These builders are referred to as \textit{HFT builders}. We move to link non-atomic arbitrage searchers and builders in this section.  

\begin{figure}
    \centering
    \includegraphics[scale=1]{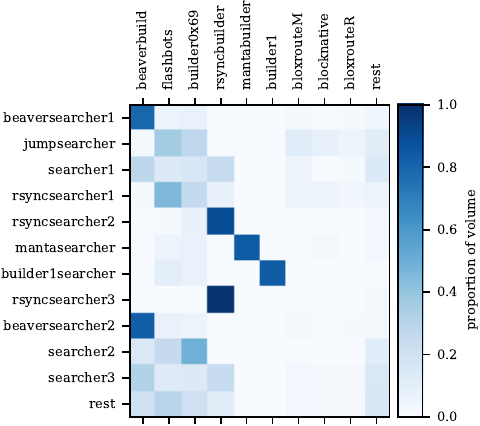}
    \caption{Proportion of non-atomic arbitrage transaction volume by each searcher (y-axis) included in blocks by each builder (x-axis). The darker a square, the stronger the relationship between a builder/searcher pair.}
    \label{fig:botsToBuilders}\vspace{-8pt}
\end{figure}

To start, we analyze what proportion of the non-atomic arbitrage volume by each of the biggest eleven searchers is included in blocks built by each of the biggest nine builders (cf. Figure~\ref{fig:botsToBuilders}). Note that we identify the biggest searchers and builders by the non-atomic arbitrage volume they traded and included in their blocks respectively. We provide a mapping from searcher/builder name to their address/public key in our data set~\cite{anonymous2023}.

The darker the tone of blue in Figure~\ref{fig:botsToBuilders}, the larger the proportion of non-atomic arbitrage volume by the searcher in blocks by the corresponding builder. We, for instance, observe that more than half (i.e., 79\% and 82\% respectively) of the non-atomic arbitrage volume from beaverseacher1 and beaverseacher2 is in blocks by beaverbuild. Thus, it is likely that these two searchers are associated with beaverbuild and we name them accordingly. Additionally, we find evidence that beaverbuild was subsiding its integrated searchers for a month as we detail in Appendix~\ref{app:beaverbuild}. This further shows that they are very likely to be the same entity.

Similarly, 83\% of the non-atomic arbitrage volume of mantasearcher is found in blocks built by the mantabuilder, while 88\% of volume from the rsyncsearcher2 and an astonishing 98\% of volume from the rsyncsearcher3 is included in blocks by the rsyncbuilder. Note that there is no link between rsyncsearcher1 and rsyncbuilder in Figure~\ref{fig:botsToBuilders} even though our naming suggests there to be a link. However, we are able to establish a link between rsyncsearcher1 and rsyncsearcher2, as the bytecode of their contracts is identical and they are thus likely to be the same entity as established in previous work~\cite{FerreiraFrontrunner2021}. We further comment that rsynbuilder was largely not operating at the same time as rsyncsearcher1~(cf. Figure~\ref{fig:rsyncbuilderCorrelation}). Finally, we find a fourth, previously not identified, builder that is likely running an integrated non-atomic arbitrage builder: builder1. To be precise, 83\% of the non-atomic arbitrage volume by builder1seacher is located in blocks built by builder1seacher.

Figure~\ref{fig:botsToBuilders} allows us to observe ties between builders and searchers. However, at the same time, we also find that non-integrated searchers (e.g, jumpsearcher, searcher1, searcher2, and seacher3) have a harder time having their non-atomic arbitrage trades included in blocks by the four builders that have their own integrated builders, i.e., their non-atomic arbitrage trades are generally in blocks by flashbots, builder0x69, etc. Thus, it seems that either other searchers are less likely to submit their transactions to these four builders with integrated searchers or the four builders favor their integrated searchers.

We further take an in-depth look at the connections between HFT builders and their integrated searchers in Appendix~\ref{app:ties}. Our analysis finds that for all builders the correlation between the share of daily blocks for which they won the PBS auction and the daily share of arbitrage volume by the respective integrated searcher is positive and the correlation is significant. Additionally, this correlation is slightly lower for beaverbuild and rsyncbuilder. This difference possibly stems from the fact that these builders also include significantly more sandwich attacks and cyclic arbitrage (i.e., common MEV types) than builder1 and mantabuilder. Thus, the value of their blocks is likely not solely driven by non-atomic arbitrage transactions.

\subsection{Impact of Non-Atomic Arbitrage Trades}

\begin{figure}[t]
    \centering
    \includegraphics[scale=1]{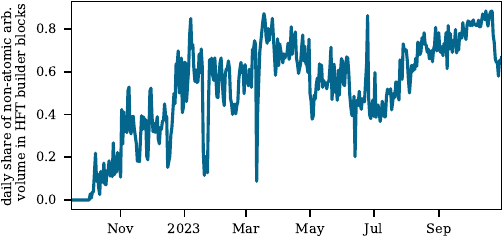}
    \caption{Daily share of non-atomic arbitrage volume in HFT builder blocks.}
    \label{fig:HFTTradesinHFTBuilderBlocks}\vspace{-6pt}
\end{figure}

To conclude our analysis, we discuss the impact of non-atomic arbitrage and the vertical integration of searchers specializing in this arbitrage. We start by noting that from 2023 onwards generally more than 50\% of non-atomic arbitrage volume is found in HFT builder blocks (cf. Figure~\ref{fig:HFTTradesinHFTBuilderBlocks}). Often this share rises to around 75\%. Recall, that there are just four HFT builders, at no time were more than three of them in operation, and from May 2023 onwards only two were still operating (cf. Figure~\ref{fig:buildersearcher}). Thus, we conclude that these builders have a strong grasp on the non-atomic arbitrage market.

\begin{figure}[b]
    \centering
    \includegraphics[scale=1]{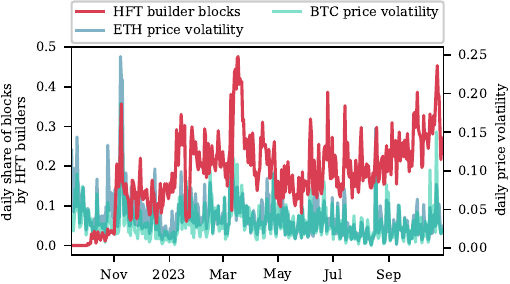}
    \caption{Daily share of blocks won by HFT builders along with the ETH and BTC price volatility.}
    \label{fig:numberHFTBuildenrBlocks}
\end{figure}

As we established in the previous section, cryptocurrency price volatility can drive non-atomic arbitrage transactions, as it signals a divergence from the price at an on-chain DEX. Thus, we analyze the relationship between ETH and BTC price volatility and the share of blocks won by HFT builders (cf. Figure~\ref{fig:numberHFTBuildenrBlocks}). We notice both an increase in the daily share of HFT builder blocks over time but also a correlation between the share of blocks won by HFT builders and the price volatility. To be precise, in October 2023 the correlation between the share of blocks built by HFT builders and the Ethereum price volatility is 0.783 with a p-value of $1.85\cdot 10^{-7}$, whereas it is 0.650 for the BTC price volatility with a p-value of $7.51\cdot 10^{-5}$. Thus, we establish that on high volatility days, these HFT builders specialized in non-atomic arbitrage transactions are more likely to win blocks. This finding is in line with a previous analysis by Gupta et al.~\cite{Gupta2023Centralizing} that showed that between April and May 2023 beaverbuild, mantabuilder, and rsyncbuilder were more likely to win the PBS auction as the price volatility of the ETH-USD market on Binance.us increases.

To further investigate the effects of cryptocurrency price volatility and thereby non-atomic arbitrage on the system, we take a closer look at the volatility in the lead-up to a block, i.e., for block $n$ the price volatility between the time block $n-1$ and block $n$ were first seen in the network.  On average (if blocks are not missed) this is twelve seconds on Ethereum. Increased volatility during this time should lead to increasingly profitable and prevalent non-atomic arbitrage transactions.

\begin{figure}[]\vspace{-0pt}
    \centering
   
    \begin{subfigure}[b]{1\linewidth}
        \includegraphics[scale=1]{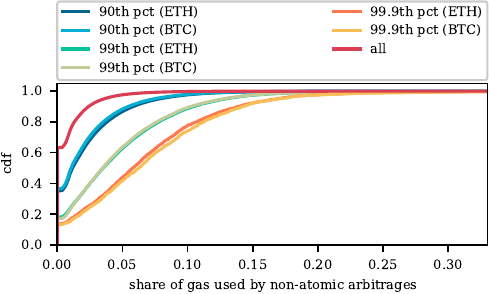}
    \caption{cdf of the share of gas used by non-atomic arbitrage}
    \label{fig:cdfGasUsage}
    \end{subfigure}\vspace{2pt}
    
    \begin{subfigure}[b]{1\linewidth}
        \includegraphics[scale=1]{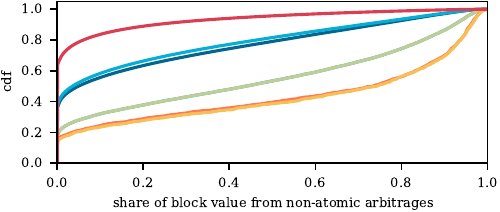}
    \caption{cdf of the share of block value from non-atomic arbitrage}
    \label{fig:cdfBlockValueFromArb}
    \end{subfigure}    

    \caption{Cumulative distribution function (cdf) of gas usage and block value of non-atomic arbitrage depending on the ETH and BTC price volatility percentile, i.e., exceeding volatility percentile indicated in the legend. The larger the volatility the larger the proportion of gas usage and block value stemming from non-atomic arbitrage. }\label{fig:cdf}\vspace{-6pt}
\end{figure}

Figure~\ref{fig:cdf} plots the cdf of gas usage and block value of non-atomic arbitrage depending on the ETH and BTC price volatility. Observe that the higher the BTC and ETH price volatility in the lead-up to the block, the larger the share of the block's gas usage by non-atomic arbitrage trades (cf. Figure~\ref{fig:cdfGasUsage}). Overall, in less than 1\% of blocks, non-atomic arbitrage transactions make up more than 5\% of the block’s total gas usage (red line). However, for blocks above the 99.9th percentile of BTC price volatility (yellow line), we see that in around 40\% of blocks, non-atomic arbitrage transactions use more than 10\% of block space, i.e., gas. Note that this is in addition to already larger blocks during times of high volatility (cf. Appendix~\ref{app:vol}). Thus, the increased prevalence of non-atomic arbitrage likely leads to increased block competition during times of high volatility. Importantly, this also leads to an increased base fee (i.e., the minimum fee required for block inclusion) for consecutive blocks as the base fee automatically increases based on the previous block size. Further, this increase in the price for transaction inclusion as well as the overfullness of the block likely leads to increased congestion in the mempool. Thus, the block space used by non-atomic arbitrage during times of high volatility impacts all Ethereum users.

We also analyze the share of block value (i.e., priority fees and direct transfers) that come from non-atomic arbitrage transactions in Figure~\ref{fig:cdfBlockValueFromArb}. Again we find that the share of value from non-atomic arbitrage transactions increases with the volatility. In fact, in around 50$\%$ of blocks where the ETH and BTC volatility is in the 99.9$^{\text{th}}$ percentile, over 80$\%$ of the block value comes from non-atomic arbitrage transactions. Note again that not only the share of value from non-atomic arbitrage but also the block value increases in times of high volatility (cf. Appendix~\ref{app:vol}). We further outline that the increased value from non-atomic arbitrage transactions in times of high volatility explains the higher likelihood that HFT builders to win blocks during times of high price volatility in ETH and BTC.

\section{Discussion}

The fees paid by non-atomic arbitrage transactions exceed current block rewards on the Ethereum PoS consensus layer repeatedly. For instance, their value exceeds the current consensus layer block reward by more than a factor of 10 in 15,360 blocks during our data collection period and we further note that their value measured in fees can be seen as a lower bound for the profit that can be extracted.

Previous works~\cite{DaianFlash2020,QinQuantifying2022} have demonstrated that MEV (i.e., high-value transactions) presents a risk to the consensus layer in PoW. To be exact, the consensus is vulnerable to time-bandit attacks, as it can be rational for the block proposer to fork the blockchain to exploit MEV in previous blocks themselves. Re-orgs, required by time-bandit attacks, have become harder in Ethereum PoS~\cite{2022reorg}, but regardless such high-value transactions present a challenge to the consensus layers. For instance, an entity controlling a significant proportion of the staking power could purposefully withhold attestations for blocks preceding its turn as block proposer to increase the chance of a possible re-org during its turn as proposer being successful. Note that the biggest staking pool currently controls around one-third of the staking power~\cite{grandjean2023ethereum} and that the losses from missed attestations are minimal.

Additionally, we show that in times of high cryptocurrency price volatility, we observe spikes in non-atomic arbitrage transactions on the Ethereum blockchain. This increase in non-atomic arbitrage transactions leads to overfull blocks resulting in increased waiting times and increased fees for the remaining Ethereum users.

\T{Centralization of Block Building.} We further comment on the centralizing effects of non-atomic arbitrage in the block construction market. Gupta et al.~\cite{Gupta2023Centralizing} have shown that HFT builders (i.e., builders with integrated non-atomic arbitrage searchers) are more likely to win blocks during times of high cryptocurrency price volatility for a one-month period in early 2023. We extend these findings to the entire history of Ethereum PoS until the end of October 2023 and highlight the relationships between builders and their integrated searchers. Our work further establishes a direct tie between the likelihood of these HFT builders winning the PBS auction and the volume of non-atomic arbitrage by their integrated searchers. As a result, during times of high cryptocurrency price volatility block building is largely left to two remaining HFT builders -- a worrying centralization of the block construction market.

\section{Mitigation}

We delve into possible mitigations to non-atomic arbitrage and its centralizing effects in the following. 

\T{Separating Top of Block.} One possible avenue to work against these centralizing effects of non-atomic arbitrage suggested by Gupta et al.~\cite{Gupta2023Centralizing} is to separate top-of-block extractions, i.e., non-atomic arbitrage that generally takes place top-of-block as these swaps wish to be the first to execute in the respective pools, and block-body extraction. By unbundling the PBS auction as outlined, the HFT builders specialized in non-atomic arbitrage would still dominate the top-of-block opportunities in times of high volatility, but would minder the effects on the rest of the block body. While we believe that this would be a step in the right direction, some problems remain. For one, the size of the top-of-block would likely have to be limited as otherwise, these top-of-block opportunities are likely to take up more than half the block space in times of high volatility. Additionally, while this approach would limit the centralizing effects of non-atomic arbitrage, it would not target the security implications (i.e., time-bandit attacks) outlined previously. 

\T{Reduction of Block Time.} A further possible mitigation is block time (i.e., time between two consecutive blocks) reduction. This reduction would lead to less profitable non-atomic arbitrage opportunities as the expected price change would be reduced as we show in Section~\ref{sec:model} and is also discussed by Milionis et al.~\cite{milionis2022automated,milionis2023automated}. If these arbitrage opportunities are less profitable their impact on the ecosystem should naturally lessen.  Note that the possible ramifications of block time reduction on the consensus layer should not be underestimated, but another possible avenue would be to move the DEX volume to Layer 2s such as Arbitrum and Optimism which already have much shorter block times. 

\section{Related Work}
\T{Maximal Extractable Value.} An active line of research is devoted to describing and measuring MEV on the Ethereum blockchain. Blockchain front-running attacks were first systematized by Eskandari et al.~\cite{EskandariSoK2020} and the term MEV was introduced by Daian et al.~\cite{DaianFlash2020} in an early description of front-running on decentralized exchanges. Our work studies a new form of MEV, i.e., non-atomic arbitrage, that was not previously investigated in depth. 

Comprehensive measurements of MEV on the Ethereum blockchain were subsequently performed by Zhou et al.~\cite{ZhouHigh2021}, Ferreira et al.~\cite{FerreiraFrontrunner2021}, and Qin et al.~\cite{QinQuantifying2022}. An in-depth study of liquidations was presented by Qin et al.~\cite{QinEmpirical2021}, whereas Wang et al.~\cite{WangCyclic2022} carried out a measurement study focused on cyclic arbitrage. In this work, we perform a measurement study of non-atomic arbitrage on DEXes, a type of MEV prevalent in Ethereum PoS, that is not included in existing MEV measurements. 

A related line of research is devoted to preventing front-running on the blockchain~\cite{bentov2019tesseract,stathakopoulou2021adding,kelkar2020order,baird2016swirlds,kursawe2020wendy,zhang2020byzantine,kelkar2021order,kelkar2021themis,cachin2021quick,reiter1994securely,miller2016honey,asayag2018fair,orda2021enforcing,zhang2022flash,constantinescu2023fair,momeni2023fairblock,tatabitovska2021mitigation,breidenbach2018enter,doweck2020multi,heimbach2022eliminating,zhou2021a2mm}, and a comprehensive overview of these approaches is provided by Heimbach et al.~\cite{HeimbachSoK2022}. Of these approaches, only those that focus on ensuring that transactions are included first in, first out are still applicable to the non-atomic arbitrage we describe.

\T{Liquidity Providers on Decentralized Exchanges.} Multiple works study the risks of liquidity providers on DEXes. Traditionally these risks have been measured in what is known as impermanent loss~\cite{Heimbach2021behavior,HeimbachRisks2022}. More recently, Milionis et al.~\cite{milionis2022automated,milionis2023automated} introduced the concept of loss-versus-rebalancing (LVR) to quantify the losses of DEX liquidity providers, which can be thought of as the expected profit of non-atomic arbitrage. Their works analytically quantify the expected profit of non-atomic arbitrage. In this work, on the other hand, we empirically study the prevalence of non-atomic arbitrage on DEXes and discuss its ramifications beyond liquidity providers.

\T{Proposer Builder Separation.} Given the immense impact of PBS on Ethereum's block construction market, multiple works~\cite{YangSoK2022,HeimbachEthereum2023,WahrstätterBlockchain2023,WahrstätterTime2023,SchwarzTime2023} investigate this novel scheme. These works investigate the block value distribution between builders and proposers, censorship resistance, and highlight the increase in centralization in the block construction market under PBS. Gupta et al.~\cite{Gupta2023Centralizing} further study HFT builders, i.e., builders known to perform non-atomic arbitrage, in the PBS scheme. Their work shows that these builders are more likely to win blocks during periods characterized by high cryptocurrency price volatility. Our work, on the other hand, studies HFT builders performing non-atomic arbitrage in detail. In comparison to this work, we take a broader view, i.e., identifying all swaps that are likely to be non-atomic arbitrage, and analyzing the prevalence of such trades, how they are distributed between searchers and builders, and their impact on the ecosystem.

\section{Conclusion}

With this work, we uncover the prevalence, centralizing effects, and implications of non-atomic arbitrage. Our measurement study highlights that more than one-fourth of DEX volume is likely to be non-atomic arbitrage and that HFT builders specialized in non-atomic arbitrage have an advantage in the PBS auction in times of high volatility. Besides highlighting the centralizing effects of non-atomic arbitrage on the block construction market, we also discuss its security implications as well as effects on the broader Ethereum ecosystem and its users. Finally, we point at possible mitigations and hope that the insights from our work positively affect future developments.



\bibliographystyle{IEEEtran}
\bibliography{bibliography}

\begin{thebibliography}{10}
\providecommand{\url}[1]{#1}
\csname url@samestyle\endcsname
\providecommand{\newblock}{\relax}
\providecommand{\bibinfo}[2]{#2}
\providecommand{\BIBentrySTDinterwordspacing}{\spaceskip=0pt\relax}
\providecommand{\BIBentryALTinterwordstretchfactor}{4}
\providecommand{\BIBentryALTinterwordspacing}{\spaceskip=\fontdimen2\font plus
\BIBentryALTinterwordstretchfactor\fontdimen3\font minus \fontdimen4\font\relax}
\providecommand{\BIBforeignlanguage}[2]{{%
\expandafter\ifx\csname l@#1\endcsname\relax
\typeout{** WARNING: IEEEtran.bst: No hyphenation pattern has been}%
\typeout{** loaded for the language `#1'. Using the pattern for}%
\typeout{** the default language instead.}%
\else
\language=\csname l@#1\endcsname
\fi
#2}}
\providecommand{\BIBdecl}{\relax}
\BIBdecl

\bibitem{WangCyclic2022}
Y.~Wang, Y.~Chen, H.~Wu, L.~Zhou, S.~Deng, and R.~Wattenhofer, ``{Cyclic Arbitrage in Decentralized Exchanges},'' in \emph{Companion Proceedings of the Web Conference 2022}, 4 2022.

\bibitem{Defillama-chain-data}
``{Total Value Locked},'' \url{https://defillama.com/chains}.

\bibitem{2022boost}
``{MEV-Boost Dashboard},'' \url{https://mevboost.pics/}.

\bibitem{Gupta2023Centralizing}
T.~Gupta, M.~M. Pai, and M.~Resnick, ``{The centralizing effects of private order flow on proposer-builder separation},'' in \emph{5th Conference on Advances in Financial Technologies}, 10 2023.

\bibitem{2022pbs}
``{Proposer-Builder Separation},'' \url{https://ethereum.org/roadmap/pbs/}.

\bibitem{DaianFlash2020}
P.~Daian, S.~Goldfeder, T.~Kell, Y.~Li, X.~Zhao, I.~Bentov, L.~Breidenbach, and A.~Juels, ``Flash boys 2.0: Frontrunning in decentralized exchanges, miner extractable value, and consensus instability,'' in \emph{2020 IEEE Symposium on Security and Privacy (SP)}, 5 2020.

\bibitem{QinEmpirical2021}
K.~Qin, L.~Zhou, P.~Gamito, P.~Jovanovic, and A.~Gervais, ``An empirical study of defi liquidations: incentives, risks, and instabilities,'' in \emph{Proceedings of the 21st ACM Internet Measurement Conference}, 11 2021.

\bibitem{Defillama-ethereum-dexes}
``{DefiLlama},'' \url{https://defillama.com/protocols/dexes/Ethereum}.

\bibitem{zeromev}
``{Zeromev},'' \url{https://zeromev.org/}.

\bibitem{mempool-guru}
{Mempool Guru}, ``{Mempool Guru},'' \url{https://mempool.guru/}, 2023.

\bibitem{YangSoK2022}
S.~Yang, F.~Zhang, K.~Huang, X.~Chen, Y.~Yang, and F.~Zhu, ``{SoK: MEV Countermeasures: Theory and Practice},'' \emph{arXiv preprint arXiv:2212.05111}, 2023.

\bibitem{binance-data}
{Binance}, ``{Historical Market Data},'' \url{https://www.binance.com/en/landing/data}, 2023.

\bibitem{CoinMarketCap2023}
\BIBentryALTinterwordspacing
``Today's cryptocurrency prices by market cap,'' 2023. [Online]. Available: \url{https://coinmarketcap.com/}
\BIBentrySTDinterwordspacing

\bibitem{adams2020uniswap}
H.~Adams, N.~Zinsmeister, and D.~Robinson, ``Uniswap v2 core,'' 2020.

\bibitem{adams2021uniswap}
H.~Adams, N.~Zinsmeister, M.~Salem, R.~Keefer, and D.~Robinson, ``Uniswap v3 core,'' 2021.

\bibitem{sushiswap-whitepaper}
{Sushiswap}, ``{Be a Crypto Chef with Sushi},'' \url{https://docs.sushi.com/pdf/whitepaper.pdf}, 2023.

\bibitem{anonymous2023}
\BIBentryALTinterwordspacing
``Non-atomic arbitrage data and code,'' 2023. [Online]. Available: \url{https://github.com/liobaheimbach/Non-Atomic-Arbitrage-in-Decentralized-Finance}
\BIBentrySTDinterwordspacing

\bibitem{QinQuantifying2022}
K.~Qin, L.~Zhou, and A.~Gervais, ``Quantifying blockchain extractable value: How dark is the forest?'' in \emph{2022 IEEE Symposium on Security and Privacy (SP)}.\hskip 1em plus 0.5em minus 0.4em\relax IEEE, 5 2022.

\bibitem{ZhouHigh2021}
L.~Zhou, K.~Qin, C.~F. Torres, D.~V. Le, and A.~Gervais, ``{High-Frequency Trading on Decentralized On-Chain Exchanges},'' in \emph{2021 IEEE Symposium on Security and Privacy (SP)}, 5 2021.

\bibitem{FerreiraFrontrunner2021}
C.~F. Torres, R.~Camino, and R.~State, ``{Frontrunner Jones and the Raiders of the Dark Forest: An Empirical Study of Frontrunning on the Ethereum Blockchain},'' in \emph{30th USENIX Security Symposium}, 8 2021.

\bibitem{HeimbachEthereum2023}
L.~Heimbach, L.~Kiffer, C.~Ferreira~Torres, and R.~Wattenhofer, ``{Ethereum's Proposer-Builder Separation: Promises and Realities},'' in \emph{2023 ACM Internet Measurement Conference (IMC), Montreal, QC, Canada}, Oct. 2023.

\bibitem{Defillama-api}
``{DefiLlama API},'' \url{https://defillama.com/docs/api}.

\bibitem{2022jump}
``{Jump Etherscan},'' \url{https://etherscan.io/address/0x9507c04b10486547584c37bcbd931b2a4fee9a41}, 2023.

\bibitem{ftx-collapse}
{Investopedia}, ``{The Collapse of FTX: What Went Wrong With the Crypto Exchange?}'' \url{https://www.investopedia.com/what-went-wrong-with-ftx-6828447}, 2023.

\bibitem{usdc-depeg}
{Cointelegraph}, ``{TUSDC depegs as Circle confirms \$3.3B stuck with Silicon Valley Bank},'' \url{https://cointelegraph.com/news/usdc-depegs-as-circle-confirms-3-3b-stuck-with-silicon-valley-bank}, 2023.

\bibitem{Blockworks-crypto-correlation}
{Blockworks}, ``{The Investor’s Guide to Crypto Correlation},'' \url{https://blockworks.co/news/the-investors-guide-to-crypto-correlation}, 2023.

\bibitem{bitcoinCorrelation}
\BIBentryALTinterwordspacing
K.~Q. Nguyen, ``The correlation between the stock market and bitcoin during covid-19 and other uncertainty periods,'' \emph{Finance Research Letters}, vol.~46, 2022. [Online]. Available: \url{https://www.sciencedirect.com/science/article/pii/S1544612321003238}
\BIBentrySTDinterwordspacing

\bibitem{2022fomc}
``Meeting calendars, statements, and minutes (2018-2024),'' \url{https://www.federalreserve.gov/monetarypolicy/fomccalendars.html}, 2023.

\bibitem{2022fed}
``How the fed impacts stocks, crypto and other investments,'' \url{https://www.bankrate.com/investing/federal-reserve-impact-on-stocks-crypto-other-investments/\#crypto}, 2023.

\bibitem{2022reorg}
G.~Konstantopoulos and V.~Buterin, ``{Ethereum Reorgs After The Merge},'' \url{https://www.paradigm.xyz/2021/07/ethereum-reorgs-after-the-merge}, 2021.

\bibitem{grandjean2023ethereum}
D.~Grandjean, L.~Heimbach, and R.~Wattenhofer, ``{Ethereum Proof-of-Stake Consensus Layer: Participation and Decentralization},'' \emph{arXiv preprint arXiv:2306.10777}, 2023.

\bibitem{milionis2022automated}
J.~Milionis, C.~C. Moallemi, T.~Roughgarden, and A.~L. Zhang, ``{Automated Market Making and Loss-Versus-Rebalancing},'' \emph{arXiv preprint arXiv:2208.06046}, 2022.

\bibitem{milionis2023automated}
J.~Milionis, C.~C. Moallemi, and T.~Roughgarden, ``{Automated Market Making and Arbitrage Profits in the Presence of Fees},'' \emph{arXiv preprint arXiv:2305.14604}, 2023.

\bibitem{EskandariSoK2020}
S.~Eskandari, S.~Moosavi, and J.~Clark, ``{SoK: Transparent Dishonesty: Front-Running Attacks on Blockchain},'' in \emph{Financial Cryptography and Data Security}, vol. 11599, 2020.

\bibitem{bentov2019tesseract}
I.~Bentov, Y.~Ji, F.~Zhang, L.~Breidenbach, P.~Daian, and A.~Juels, ``Tesseract: Real-time cryptocurrency exchange using trusted hardware,'' in \emph{Proceedings of the 2019 ACM SIGSAC Conference on Computer and Communications Security}, ser. CCS '19, 2019.

\bibitem{stathakopoulou2021adding}
C.~Stathakopoulou, S.~R{\"u}sch, M.~Brandenburger, and M.~Vukoli{\'c}, ``{Adding Fairness to Order: Preventing Front-Running Attacks in BFT Protocols using TEEs},'' in \emph{2021 40th International Symposium on Reliable Distributed Systems (SRDS)}, 2021.

\bibitem{kelkar2020order}
M.~Kelkar, F.~Zhang, S.~Goldfeder, and A.~Juels, ``{Order-Fairness for Byzantine Consensus},'' in \emph{Annual International Cryptology Conference}, 2020.

\bibitem{baird2016swirlds}
L.~Baird, ``{The Swirlds Hashgraph Consensus Algorithm: Fair, Fast, Byzantine Fault Tolerance},'' \emph{Swirlds Tech Reports SWIRLDS-TR-2016-01, Tech. Rep}, 2016.

\bibitem{kursawe2020wendy}
K.~Kursawe, ``{Wendy, the Good Little Fairness Widget: Achieving Order Fairness for Blockchains},'' in \emph{2nd ACM Conference on Advances in Financial Technologies}, 2020.

\bibitem{zhang2020byzantine}
Y.~Zhang, S.~Setty, Q.~Chen, L.~Zhou, and L.~Alvisi, ``{Byzantine Ordered Consensus Without Byzantine Oligarchy},'' in \emph{14th $\{$USENIX$\}$ Symposium on Operating Systems Design and Implementation ($\{$OSDI$\}$ 20)}, 2020.

\bibitem{kelkar2021order}
M.~Kelkar, S.~Deb, and S.~Kannan, ``{Order-Fair Consensus in the Permissionless Setting},'' \emph{IACR Cryptol. ePrint Arch.}, 2021.

\bibitem{kelkar2021themis}
M.~Kelkar, S.~Deb, S.~Long, A.~Juels, and S.~Kannan, ``{Themis: Fast, Strong Order-Fairness in Byzantine Consensus},'' Cryptology ePrint Archive, Report 2021/1465, 2021.

\bibitem{cachin2021quick}
C.~Cachin, J.~Mi{\'c}i{\'c}, and N.~Steinhauer, ``{Quick Order Fairness},'' in \emph{Financial Cryptography and Data Security (FC), Grenada}, 2022.

\bibitem{reiter1994securely}
M.~K. Reiter and K.~P. Birman, ``{How to Securely Replicate Services},'' \emph{ACM Transactions on Programming Languages and Systems (TOPLAS)}, vol.~16, no.~3, 1994.

\bibitem{miller2016honey}
A.~Miller, Y.~Xia, K.~Croman, E.~Shi, and D.~Song, ``{The Honey Badger of BFT Protocols},'' ser. CCS '16, 2016.

\bibitem{asayag2018fair}
A.~Asayag, G.~Cohen, I.~Grayevsky, M.~Leshkowitz, O.~Rottenstreich, R.~Tamari, and D.~Yakira, ``{A Fair Consensus Protocol for Transaction Ordering},'' in \emph{2018 IEEE 26th International Conference on Network Protocols (ICNP)}, 2018.

\bibitem{orda2021enforcing}
A.~Orda and O.~Rottenstreich, ``{Enforcing Fairness in Blockchain Transaction Ordering},'' \emph{Peer-to-peer Networking and Applications}, vol.~14, no.~6, 2021.

\bibitem{zhang2022flash}
H.~Zhang, L.-H. Merino, V.~Estrada-Galinanes, and B.~Ford, ``{Flash freezing flash boys: Countering blockchain front-running},'' in \emph{2022 IEEE 42nd International Conference on Distributed Computing Systems Workshops (ICDCSW)}, 2022.

\bibitem{constantinescu2023fair}
A.~Constantinescu, D.~Ghinea, L.~Heimbach, Z.~Wang, and R.~Wattenhofer, ``{A Fair and Resilient Decentralized Clock Network for Transaction Ordering},'' in \emph{27th International Conference on Principles of Distributed Systems (OPODIS), Tokyo, Japan}, Dec. 2023.

\bibitem{momeni2023fairblock}
P.~Momeni, S.~Gorbunov, and B.~Zhang, ``{Fairblock: Preventing Blockchain Front-Running with Minimal Overheads},'' in \emph{Security and Privacy in Communication Networks: 18th EAI International Conference, SecureComm 2022, Virtual Event, October 2022, Proceedings}, 2023.

\bibitem{tatabitovska2021mitigation}
A.~Tatabitovska, O.~Ersoy, and Z.~Erkin, ``{Mitigation of Transaction Manipulation Attacks in UniSwap},'' 2021.

\bibitem{breidenbach2018enter}
L.~Breidenbach, P.~Daian, F.~Tram{\`e}r, and A.~Juels, ``{Enter the Hydra: Towards Principled Bug Bounties and Exploit-Resistant Smart Contracts},'' in \emph{27th USENIX Security Symposium}, 2018.

\bibitem{doweck2020multi}
Y.~Doweck and I.~Eyal, ``{Multi-Party Timed Commitments},'' \emph{arXiv preprint arXiv:2005.04883}, 2020.

\bibitem{heimbach2022eliminating}
L.~Heimbach and R.~Wattenhofer, ``{Eliminating Sandwich Attacks with the Help of Game Theory},'' in \emph{ACM Asia Conference on Computer and Communications Security (ASIA CCS), Nagasaki, Japan}, Jun. 2022.

\bibitem{zhou2021a2mm}
L.~Zhou, K.~Qin, and A.~Gervais, ``{A2MM: Mitigating Frontrunning, Transaction Reordering and Consensus Instability in Decentralized Exchanges},'' 2021.

\bibitem{HeimbachSoK2022}
L.~Heimbach and R.~Wattenhofer, ``{SoK: Preventing Transaction Reordering Manipulations in Decentralized Finance},'' in \emph{4th ACM Conference on Advances in Financial Technologies (AFT), Cambridge, MA, USA}, Sep. 2022.

\bibitem{Heimbach2021behavior}
L.~Heimbach, Y.~Wang, and R.~Wattenhofer, ``{Behavior of Liquidity Providers in Decentralized Exchanges},'' in \emph{2021 Crypto Valley Conference on Blockchain Technology (CVCBT), Rotkreuz, Switzerland}, Oct. 2021.

\bibitem{HeimbachRisks2022}
L.~Heimbach, E.~Schertenleib, and R.~Wattenhofer, ``{Risks and Returns of Uniswap V3 Liquidity Providers},'' in \emph{4th ACM Conference on Advances in Financial Technologies (AFT), Cambridge, MA, USA}, Sep. 2022.

\bibitem{WahrstätterBlockchain2023}
A.~Wahrstätter, J.~Ernstberger, A.~Yaish, L.~Zhou, K.~Qin, T.~Tsuchiya, S.~Steinhorst, D.~Svetinovic, N.~Christin, M.~Barczentewicz, and A.~Gervais, ``{Blockchain Censorship},'' \emph{arXiv preprint arXiv:2305.18545}, 2023.

\bibitem{WahrstätterTime2023}
A.~Wahrstätter, L.~Zhou, K.~Qin, D.~Svetinovic, and A.~Gervais, ``{Time to Bribe: Measuring Block Construction Market},'' \emph{arXiv preprint arXiv:2305.16468}, 2023.

\bibitem{SchwarzTime2023}
C.~Schwarz-Schilling, F.~Saleh, T.~Thiery, J.~Pan, N.~Shah, and B.~Monnot, ``{Time is Money: Strategic Timing Games in Proof-of-Stake Protocols},'' \emph{arXiv preprint arXiv:2305.09032}, 2023.

\end{thebibliography}
%

\appendices

\section{HFT Builders}

\subsection{Searcher - Builder Ties}\label{app:ties}

\begin{figure}[h!]\vspace{-6pt}
    \centering
    \begin{subfigure}[b]{1\linewidth}
        \includegraphics[scale=1]{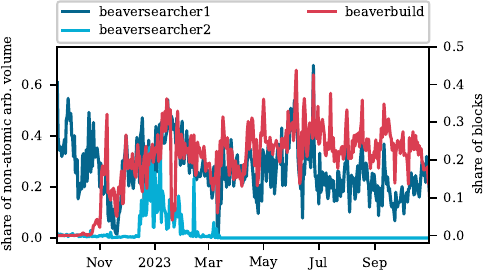}
    \caption{beaverbuild}\vspace{2pt}
    \label{fig:beaverbuildCorrelation}
    \end{subfigure}
    
    \begin{subfigure}[b]{1\linewidth}
        \includegraphics[scale=1]{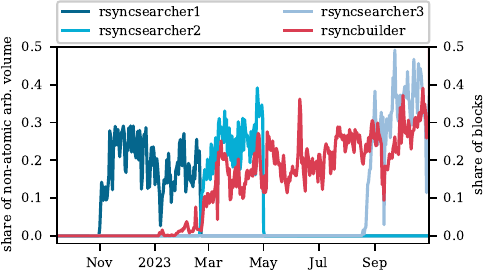}
    \caption{rsyncbuilder}\vspace{2pt}
    \label{fig:rsyncbuilderCorrelation}
    \end{subfigure}    
    
    \begin{subfigure}[b]{1\linewidth}
        \includegraphics[scale=1]{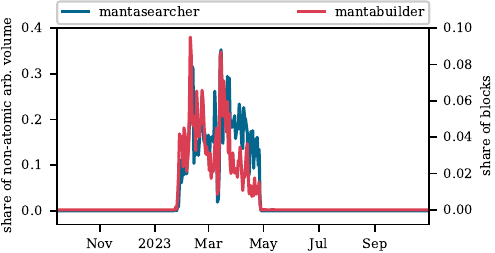}
    \caption{mantabuilder}\vspace{2pt}
    \label{fig:mantabuilderCorrelation}
    \end{subfigure}
    
    \begin{subfigure}[b]{1\linewidth}
        \includegraphics[scale=1]{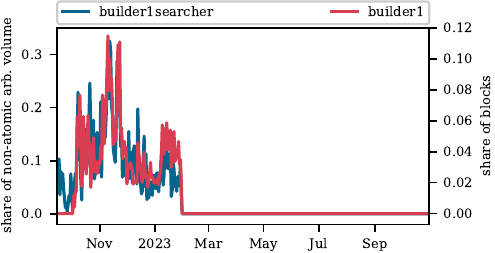}
    \caption{builder1}\vspace{2pt}
    \label{fig:builder1Correlation}
    \end{subfigure}   
    \caption{We plot the daily share of the non-atomic arbitrage volume of the searchers and the share of blocks built by the builder.}\label{fig:buildersearcher}\vspace{-12pt}
\end{figure}
In the following, we provide an in-depth analysis of the ties between HFT builders and integrated searchers.

In Figure~\ref{fig:buildersearcher} we plot the daily share of blocks on the Ethereum network they built against the daily share of the non-atomic arbitrage volume executed by their integrated searchers for each HFT builder. Interestingly, both beaverbuild (cf. Figure~\ref{fig:beaverbuildCorrelation}) and rsyncbuilder (cf. Figure~\ref{fig:rsyncbuilderCorrelation}) were active as searchers, before their builders started to be active. For beaverbuild the delay is around one and a half months whereas it is around two months for rsyncbuilder. Additionally, there was a period where the rsyncbuilder was operating from May to September 2023 but we could associate no searcher with rsyncbuilder during that time window. However, during times when at least one integrated searcher and the builders were operating we find that there is a correlation between the share of non-atomic arbitrage volume swapped by the integrated searchers and the share of blocks won by the respective builders. We find that the correlation between the share of non-atomic arbitrage volume by beaversearcher1 and beaversearcher2 and the share of block won by beaverbuild is 0.244 with a p-value of $1.83\cdot 10^{-6}$ -- indicating very high statistical significance even though the correlation is relatively small. During the operation of rsynsearcher2, the correlation between the share of non-atomic arbitrage volume by rsyncseracher2 and the share of block won by rsyncbuilder is 0.805 with a p-value of $5.68\cdot 10^{-18}$, whereas it correlation is 0.502 with a p-value of $4.50\cdot 10^{-6}$ during lifetime of rsyncseacher3. 

Turing to mantabuilder (cf. Figure~\ref{fig:mantabuilderCorrelation})  and builder1 (cf. Figure~\ref{fig:builder1Correlation}), we observe for both of them the searcher and builder are operating for the same period. During that time the correlation between the share of volume by the searcher and the share of blocks won by the builder is higher than 0.7. We further note that interestingly, mantabuilder comes into operation with its searcher just as builder1 closes operation.

\begin{table}[t]\vspace{-6pt}
\scriptsize

\centering

\begin{adjustbox}{width=\linewidth}
\begin{tabular}{@{}lrrrr@{}}
\toprule
 & beaverbuild & rsyncbuilder & builder1 & mantabuilder \\
\midrule
sandwich attacks per block & 0.041 & 0.043 & 0.000 & 0.001 \\
cyclic arbitrage per block & 0.008 & 0.006 & 0.003 & 0.001 \\
\bottomrule
\end{tabular}

\end{adjustbox}

    \caption{Sandwich attack and cyclic arbitrage transactions included in blocks by the four builders identified to have integrated non-atomic arbitrage searchers. }
    \label{tab:fourbuilder}\vspace{-8pt}
\end{table} 

To explain the lower correlation between the market shares of beaverbuild and its searchers in comparison to the other HFT builders, we take a look at the number of sandwich attacks and cyclic arbitrage transactions these builders include in their blocks. In Table~\ref{tab:fourbuilder} find that beaverbuild and rsyncbuilder, where we also saw a lower correlation during the operation of rsyncseacher3, are much more likely to include sandwich attacks and cyclic arbitrage in their blocks, than builder1 and mantabuilder. Thus, we believe that the block value of beaverbuild and rsyncbuilder blocks also come from other types of MEV. Therefore, it is not only the non-atomic arbitrage volume that determines the value of their blocks even though there is a significant correlation.

\subsection{Beaverbuild}\label{app:beaverbuild}

\begin{figure}[h]\vspace{-4pt}
    \centering
    \includegraphics[scale=1]{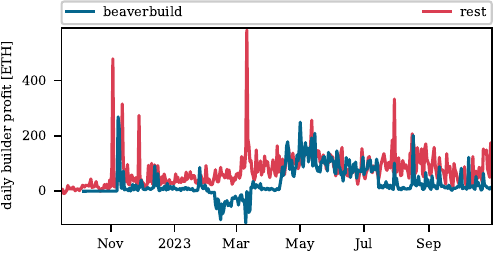}
    \caption{Daily builder profit (i.e., priority fees and coinbase transfers minus the payment to the proposer) of beaverbuild and the remaining builders. Notice that for a month the beaverbuild profit was negative. }
    \label{fig:builderProfit}\vspace{-4pt}
\end{figure}

Next, we further cement the relationship between beaverbuild and its integrated searchers by highlighting that beaverbuild was likely subsidizing its searchers for around a month. Figure \ref{fig:builderProfit} shows beaverbuild's daily profit (i.e., the difference between priority fees and direct transfers from transactions and the payment to the proposer) and we observe that it is consistently negative between 3 February and 14 March 2023 (from block 16,544,549 to 16,822,491). Thus, during this time beaverbuild was offering proposers a higher bid than what transactions in the blocks have paid in fees. For example, in block 16,627,349 these fees amounted to 56.18 ETH even though the total amount received in fees was 0.059 ETH. Hence, it is very likely that beaverbuild has another source of income, whereby they can afford to pay the proposers such large fees. We note that these suspicious negative profits were also identified by Heimbach et al.~\cite{HeimbachEthereum2023}, but no further explanation was given. 

We find that the beaverbuild was in all likelihood subsidizing its own builder. In the following, we check whether the large fees could be explained by the prevalence of these searchers within blocks that offer a higher bid. For this, we look at all blocks that were built by beaverbuild during this time, and which blocks offered larger bids than what they earned in transaction fees.

During this time beaverbuild built 48,606 blocks, with 35,584 (73.2\%) blocks offering the proposer a value larger than the fee that they received.  We find that 34,682 blocks (around 97.46\% of blocks) had at least one transaction from either beaversearcher1 or beaversearcher2. Furthermore, in the remaining 13,022 blocks only 2,802 blocks (around 21.5\%) had at least one transaction from either searcher. During this time beaverbuild received a total of 1,941.1 ETH in fees but paid proposers 6,620.94 ETH. This corresponds to a loss of over 4,679.84 ETH. This is further evidence of integrated builder searchers and points to the fact that these searchers are run by beaverbuild themselves. Furthermore, this makes our claim stronger that builders themselves profit from non-atomic arbitrage since beaverbuild had a loss of over 3,195.76 ETH on-chain. For all the non-beaverbuild blocks over this time, we find that out of 211,960 only 17,378 blocks (7.6\%) offer bids to proposers that are larger than what they received in fees from transactions. There was a total of 31,954.07 ETH paid in fees by transactions, and 28,245.00 ETH being paid to proposers. 

\section{Effects of Volatility}\label{app:vol}

\begin{figure}[t]\vspace{-0pt}
    \centering
   
    \begin{subfigure}[b]{1\linewidth}
        \includegraphics[scale=1]{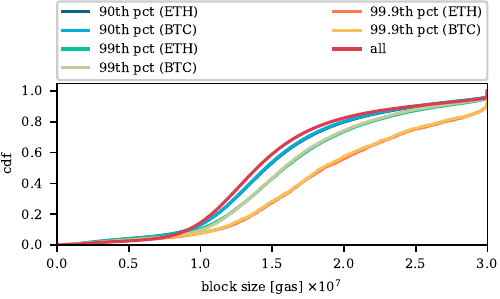}\vspace{-4pt}
    \caption{cdf of the block size}
    \label{fig:cdfGasUsageBlock}
    \end{subfigure}\vspace{2pt}
    
    \begin{subfigure}[b]{1\linewidth}
        \includegraphics[scale=1]{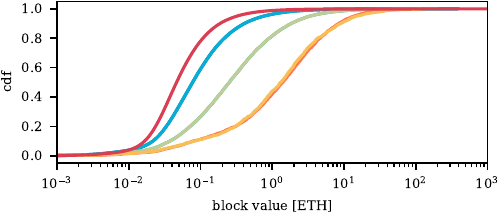}\vspace{-4pt}
    \caption{cdf of block value}
    \label{fig:cdfBlockValue}
    \end{subfigure}    

    \caption{Cumulative distribution function (cdf) of gas usage and block value depending on the ETH and BTC price volatility percentile, i.e., exceeding volatility percentile indicated in the legend.}\label{fig:cdfapp}\vspace{-6pt}
\end{figure}

We analyze the effects of high cryptocurrency price volatility on the block sizes and block value (i.e., priority fees and direct transfers) in Figure~\ref{fig:cdfapp}. First, we note blocks with high cryptocurrency price volatility tend to be fuller as we see in Figure~\ref{fig:cdfGasUsageBlock}. Note that in Ethereum the target size for a block is $1.5\cdot 10^7$, while the maximum block size is  $3\cdot 10^7$. For blocks in the 99.9$^{\text{th}}$ percentile in terms of price volatility only around 20\% of them are at target size or smaller, whereas around 50\% of overall blocks are no larger than target size.

Additionally, we find that blocks built during periods of high volatility are more valuable (cf. Figure~\ref{fig:cdfBlockValue}). For instance for blocks in the 99.9$^{\text{th}}$ percentile in terms of price volatility, more than half of them are at least worth 1 ETH, in comparison to less than 1\% overall.

\begin{table*}[b]\vspace{-6pt}
\scriptsize

\centering

\begin{adjustbox}{width=\linewidth}

\begin{tabular}{@{}lrcccccccc@{}}
\toprule
 &  total swaps & simple   & private &first swap in pool  & top tokens  & coinbase transfer & priority fee &coinbase transfer or priority fee  & all \\
& & heuristic 1 & heuristic 2 & heuristic 4&heuristic 5 & & &heuristic 3\\
\midrule
\textbf{beaversearchers} & 1,188,696 & 0.982 & 0.815 & 0.826 & 0.822 & 0.608 & 0.313 & 0.917 & 0.580 \\
\textbf{rsyncsearchers} & 346,570 & 1.000 & 0.997 & 0.924 & 0.931 & 0.059 & 0.895 & 0.954 & 0.884 \\
jumpsearcher & 232,226 & 1.000 & 0.956 & 0.986 & 0.989 & 0.000 & 1.000 & 1.000 & 0.943 \\
builder1searcher & 86,709 & 0.999 & 0.881 & 0.895 & 0.926 & 0.853 & 0.138 & 0.989 & 0.805 \\
\textbf{mantasearcher} & 60,813 & 0.999 & 0.941 & 0.921 & 0.955 & 0.940 & 0.057 & 0.977 & 0.879 \\
searcher1 & 57,824 & 0.978 & 0.996 & 0.990 & 0.993 & 0.000 & 1.000 & 1.000 & 0.963 \\
searcher2 & 46,878 & 0.441 & 0.999 & 0.700 & 0.479 & 0.075 & 0.424 & 0.499 & 0.174 \\
searcher3 & 16,201 & 1.000 & 0.998 & 0.989 & 1.000 & 1.000 & 0.000 & 1.000 & 0.988 \\
overall & 77,019,583 & 0.651 & 0.317 & 0.280 & 0.151 & 0.036 & 0.568 & 0.601 & 0.028 \\
\bottomrule
\end{tabular}
\end{adjustbox}

    \caption{Proportion of transactions from large non-atomic arbitrageurs our five heuristics apply to individually and together in comparison to all swaps on DEXes. The previously identified integrated searchers are highlighted in bold.}
    \label{tab:heuristic1}\vspace{-6pt}
\end{table*} 

\section{MEV}\label{app:mev}
\begin{figure}[t]\vspace{-0pt}
    \centering
   
    \begin{subfigure}[b]{1\linewidth}
        \includegraphics[scale=1]{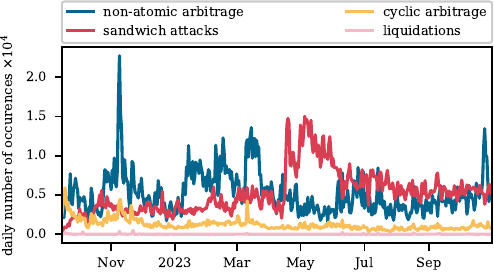}\vspace{-4pt}
    \caption{Daily number of occurrences of non-atomic arbitrage in comparison to sandwich attacks, cyclic arbitrage and liquidations.}
    \label{fig:MEVtypes}
    \end{subfigure}\vspace{4pt}
    
    \begin{subfigure}[b]{1\linewidth}
        \includegraphics[scale=1]{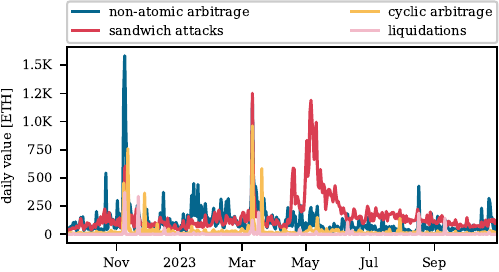}\vspace{-4pt}
    \caption{Daily value of of non-atomic arbitrage in comparison to sandwich attacks, cyclic arbitrage and liquidations.}
    \label{fig:MEVtypespayment}
    \end{subfigure}    

    \caption{Comparison between non-atomic arbitrage and well-studied types of MEV, i.e., sandwich attacks, cyclic arbitrage, and liquidations.}\label{fig:mev}\vspace{-6pt}
\end{figure}

In Figure~\ref{fig:mev}, we provide a comparison between non-atomic arbitrage and well-studied types of MEV, i.e., sandwich attacks, cyclic arbitrage, and liquidations. First, we compare the daily number of occurrences of each type of MEV in Figure~\ref{fig:MEVtypes}. Note that for sandwich attacks, which consist of a frontrunning transaction, the victim's transaction, and a backrunning transaction, we count this set of transactions as a single transaction. Non-atomic arbitrage and sandwich attacks dominate MEV in terms of a number of occurrences. Additionally, in the first part of our data collection window, the number of occurrences of non-atomic arbitrage exceeds that of sandwich attacks. Then around May 2023, it switches and there are more sandwich attacks than occurrences of non-atomic arbitrage. Notably, this coincides with \texttt{jaredfromsubway.eth}, i.e., a sandwich attack bot, entering the MEV market. By July 2023, the occurrences of non-atomic arbitrage and sandwich attacks are quite similar and continue to be so until the end of our data collection window. 

We further compare the value, i.e., the fees paid to the proposer/builder through priority fees and coinbase transfers, of the four types of MEV (cf. Figure~\ref{fig:MEVtypespayment}). Again we notice that non-atomic arbitrage and sandwich attacks are responsible for the majority of the value. To put it in numbers, the average daily non-atomic arbitrage value is 112 ETH, whereas it is 161 ETH for sandwich attacks, 26 ETH for cyclic arbitrage, and 5 ETH for liquidations. We again see an increase in the value of sandwich attacks, as \texttt{jaredfromsubway.eth} enters the MEV market around May 2023, which then declines again. Additionally, we observe that on days with high cryptocurrency price volatility, such as the FTX collapse in November 2022 or the USDC depeg in March 2023, we observe spikes in all types of MEV. 

To conclude, the scale of non-atomic arbitrage in terms of number of occurrences and in terms of value is comparable to that of sandwich attacks and significantly exceeds that of cyclic arbitrage and liquidations. 
\balance

\section{Heuristics}\label{app:heuristics}

To further ensure the validity of our heuristics, we investigate the portion of transactions we flag as non-atomic arbitrages in Table~\ref{tab:heuristic1} for each of the largest eleven searchers that we identified. The previously identified non-atomic arbitrage searchers that we studied to develop these heuristics are included for completeness and marked in bold.  These searchers were chosen as Gupta et al. \cite{Gupta2023Centralizing} previously showed that these searchers were likely involved in CEX-DEX arbitrages. We further note that these searchers are implemented as smart contracts and are likely to specialize in one profitable type of transaction. Thus, if we identify the majority of a searcher's transactions as non-atomic arbitrage it is very likely that this classification is accurate. Turning to Table~\ref{tab:heuristic1}, we notice that for all newly identified searcher except one (i.e., seacher2), more than 80\% of their transactions fulfill all our five heuristics. This leads us to conclude that these searchers are in fact performing non-atomic arbitrage. We believe that we are more likely to undercount non-atomic arbitrage as the remaining swaps from these wallets are also likely to be (failed) non-atomic arbitrage swaps (i.e., they tend to be simple transfers and the first in the pool). However, we do not include these in our analysis to avoid overstating the volume of non-atomic arbitrage. Finally, we note that for searcher2 our heuristics only apply to 17\% of swaps on DEXes.  While the swaps we identify as non-atomic arbitrage appear to be simple swaps, this searcher also appears to perform other types of high-value transactions given that less than half of its swaps classify as simple.

\end{document}